\documentclass[aps, pra, reprint, tightenlines, letterpaper, amsmath, amssymb, preprintnumbers, floatfix, longbibliography, nofootinbib,superscriptaddress]{revtex4-2}

\usepackage{tabularx}
\usepackage{dsfont}
\usepackage{extarrows}
\usepackage{amsmath}
\usepackage{amssymb}
\usepackage{physics}
\usepackage{tabu}
\usepackage{tabularx}
\usepackage{bm}
\usepackage{graphicx}
\usepackage{placeins}
\usepackage{textgreek} 
\usepackage{multirow}
\usepackage{makecell}
\usepackage{soul}
\usepackage{diagbox}
\usepackage{multirow}
\usepackage{listings}
\usepackage{graphicx}
\usepackage[table,xcdraw]{xcolor}

\usepackage[hypertexnames=false]{hyperref}
\hypersetup{
    colorlinks=true,       
    linkcolor=blue,          
    citecolor=blue,        
    filecolor=blue,      
    urlcolor=blue           
}

\usepackage[normalem]{ulem}
\newcolumntype{Y}{>{\centering\arraybackslash}X}


\usepackage{orcidlink}

\makeatletter  
\renewcommand\onecolumngrid{
\do@columngrid{one}{\@ne}%
\def\set@footnotewidth{\onecolumngrid}
\def\footnoterule{\kern-6pt\hrule width 1.5in\kern6pt}%
}
\makeatother    
\begin{document}

\title{Studying energy-resolved transport with wavepacket dynamics on quantum computers}

\author{Melody Lee\,\orcidlink{0009-0009-7833-8902}}
\email{mlee769@gatech.edu}
\affiliation{College of Computing, Georgia Institute of Technology}
\affiliation{Institute for Quantum Information and Matter, California Institute of Technology}
\author{Roland C. Farrell\,\orcidlink{0000-0001-7189-0424}}
\email{rolandf@caltech.edu}
\affiliation{Institute for Quantum Information and Matter, California Institute of Technology}
\affiliation{Department of Physics, California Institute of Technology}

\date{\today}

\begin{abstract}
\noindent
Probing energy-dependent transport in quantum simulators requires preparing states with tunable energy and small energy variance.
Existing approaches often study quench dynamics of simple initial states, such as computational basis states, which are far from energy eigenstates and therefore limit the achievable energy resolution.
In this work, we propose using wavepackets to probe transport properties with improved energy resolution.
To demonstrate the utility of this approach, we prepare and evolve wavepackets on Quantinuum’s {\tt H2-2} quantum computer and identify an energy-dependent localization transition in the Anderson model on an $8\times7$ lattice--a finite-size mobility edge. 
We observe that a wavepacket initialized at low energy remains spatially localized under time evolution, while a high-energy wavepacket delocalizes, consistent with the presence of a mobility edge. 
Crucial to our experiments is an error mitigation strategy that infers the noiseless output bit string distribution using maximum-likelihood estimation.
Compared to post-selection, this method removes systematic errors and reduces statistical uncertainty by up to a factor of $5$.
We extend our methods to the many-particle regime by developing a quantum algorithm for preparing quasiparticle wavepackets in a one-dimensional model of interacting fermions.
This technique has modest quantum resource requirements, making wavepacket-based studies of transport in many-body systems a promising application for near-term quantum computers.
\end{abstract}

\maketitle

\section{Introduction}
\noindent
Understanding charge and energy transport from first principles is central to predicting the electrical and thermal conductivities of quantum systems. 
A key challenge is determining how transport depends on energy, 
since different energy scales can exhibit qualitatively distinct behavior. 
A prominent example of this is mobility edges, characterized by a critical energy that separates localized eigenstates—whose wavefunctions are confined to a finite spatial region—from extended eigenstates that are spread across the entire system. 
Mobility edges can arise in disordered systems, where spatial inhomogeneities introduced by random impurities, external fields, or lattice defects fundamentally alter transport. 
Such disorder-induced mobility edges provided an early framework for understanding certain metal-insulator phase transitions.~\cite{anderson1958absence,Billy_2008,PhysRevLett.114.146601}. 
This phenomenon can often be studied within simple lattice models, where the effects of disorder, interactions, and energy can be systematically isolated and analyzed.

One way of studying these models is by emulating them using a quantum system that can be manipulated in experiment.
In ultracold-atom platforms, synthetic one-dimensional lattices have been engineered and used to identify single-particle mobility edges in the presence of quasi-periodic disorder~\cite{An:2018qfi,Luschen:2017cir,PhysRevLett.126.040603}.
These experiments are able to measure the particle-density in the ground state and most-excited state, and identify differences due to a mobility edge.
However, limitations in state preparation limit the energy resolution that can be achieved, and preclude a quantification of the critical energy.

Beyond single-particle systems, experiments using superconducting qubit arrays have begun to explore many-body 
mobility edges in lattices with random disorder~\cite{Guo:2019eke,Gong:2020dav}.
This is an exciting direction as studying many-body localization at scale is likely intractable with classical computing~\cite{Sierant:2024khi,Choi_2016,yao2023observation,li2025many}.
These experiments time-evolve Fock states (computational basis states) under the native device Hamiltonian, with the spread of local observables distinguishing between a localized and extended phase.
By computing $\langle H\rangle$ of the initial state, and time-evolving multiple initial states, Ref.~\cite{Guo:2019eke} charted the energy dependence of the many-body localization phase diagram.
The agreement between experiment and numerics is remarkable.
However, limitations in state preparation restrict the achievable energy resolution.
This limitation is expected as, for generic local Hamiltonians, the energy variance of a product state scales with system size $N$ as $(\Delta E)^2 = \langle H^2\rangle-\langle H\rangle^2 = \mathcal{O}(N)$.

The preparation of initial states with a lower energy variance across the spectrum often requires fine-grained experimental control that is not readily available on analog quantum simulators.
In this work, we propose using digital quantum computers to investigate energy-resolved transport by simulating wavepacket dynamics.
The wavepackets we consider are superpositions of low-energy excitations that form a gaussian-shaped ``lump" of energy in a target spatial region.
Time evolving this initial state provides insight into how energy and other local charges are transported throughout the system.  
Wavepackets feature a tunable momentum, which, through the dispersion relation, allows for a range of energies to be accessed. 
For small perturbations to a translationally invariant Hamiltonian by, e.g. disorder, the dominant contribution to the energy variance scales as $(\Delta E)^2 = \mathcal{O}(N^{-2/D})$ where $D$ is the spatial dimension.
Wavepackets therefore offer a large improvement in energy resolution over product states, particularly in lower dimensions.

We apply this approach to identify a finite-size mobility edge in the 2D Anderson model from the time-evolution of single-particle wavepackets with different energies.
Due to the mobility edge, low-energy wavepackets remain localized near their initial position, whereas high-energy wavepackets disperse across the lattice.
This is demonstrated on a $8\times 7$ lattice using Quantinuum's {\tt H2-2} trapped-ion quantum computer. 
Excellent agreement between classical and quantum simulations of the wavepacket evolution is achieved after the application of error mitigation.
Our error mitigation strategy uses maximum-likelihood estimation (MLE) to reconstruct the most probable noiseless output distribution assuming identical and independently distributed (IID) bit-flip errors.
Compared to symmetry-based post-selection, this error mitigation strategy removes systematic errors while simultaneously shrinking statistical error bars by up to $5\times$.
Similar approaches for mitigating bit-flip errors using MLE were recently developed in Refs.~\cite{baron2024qubit,Chandramouli:2025rfx}.
After applying MLE error mitigation, we are able to extract a clear separation between the time-evolved spreading of a low-energy (localized) and high-energy (extended) wavepacket that is due to the mobility edge.

These quantum simulations were in the Anderson model with a single particle.
Realizing the utility of quantum computers requires extending to the many-particle regime where ab-initio simulations at scale are generally intractable using classical computing~\cite{Feynman1982,Feynman1986,Troyer_2005,Lloyd1073, Benioff1980,Dalzell:2023ywa}.
As a step in this direction, we develop a quantum algorithm for preparing wavepackets in a one-dimensional model of interacting spinless fermions (the $XXZ$ model).
Our method combines the variational preparation of the $XXZ$ ground state from  Ref.~\cite{Yu:2022ivm} with the use of W states~\cite{Dur:2000zz} for efficient wavepacket preparation from Ref.~\cite{Farrell:2025nkx}, see also Ref.~\cite{Bennewitz:2025nhz}. 
The wavepackets that we prepare are superpositions of low-energy excitations, i.e. composed of single quasiparticle excitations of the many-body system.
The dynamics of these wavepackets thus probe transport properties over an energy interval set by the quasiparticle bandwidth, which is independent of system size.
As a result, this approach does not address transport at fixed energy density, as would be relevant to many-body localization~\cite{Alet_2018}. 
Instead, it is primarily applicable to single quasiparticle transport.

This paper is organized as follows: Section~\ref{s:preliminaries} presents classical simulations of single-particle wavepacket dynamics in the 2D Anderson model.
It is shown that wavepackets initialized at different momenta exhibit qualitatively different spreading, revealing the presence of a finite-size mobility edge.
Section~\ref{s:MLE} introduces the MLE error mitigation strategy and Section~\ref{s:simulations} presents the results of quantum simulations of wavepacket dynamics using {\tt H2-2}. 
In Section~\ref{s:halffilling}, a quantum algorithm for preparing wavepackets in the one-dimensional $XXZ$ model is developed, and the results of classical simulations verifying the method are presented. Section~\ref{s:discussion} summarizes and discusses future directions.

\section{Detecting finite-size mobility edges in the 2D Anderson model with wavepacket dynamics}\label{s:preliminaries}
\noindent 
Single-particle eigenstates in a translationally invariant system are plane waves whose probability density is extended across the whole volume.
In many realistic systems, translational invariance is broken by some kind of disorder that disrupts the periodic lattice structure.
In one and two dimensions, a non-zero density of disorder leads to localized energy eigenstates, whereas in three dimensions there is a critical disorder strength above which the eigenstates become localized~\cite{anderson1958absence,PhysRevLett.42.673,RevModPhys.57.287}.
Wavefunctions localized around a position $\vec{x}_0$ have the form $|\psi_E\rangle \propto \exp[-|{\vec x} - {\vec x}_0|/\xi(E)]$ where $\xi(E)$ is the energy-dependent localization length.
This section focuses on two dimensions, where in the infinite-volume $N\to\infty$ limit, all eigenstates are localized with $\xi(E)/N \to 0$.
However, for weak disorder and in a finite-size system, the localization length can be exponentially larger than the side lengths of the simulation volume~\cite{PhysRevLett.42.673,RevModPhys.57.287}.
This leads to a finite-size mobility edge that separates localized and extended energy eigenstates.
This is discussed further in Appendix~\ref{a:qsim_params}.

\begin{figure*}
    \centering
    \includegraphics[width=\linewidth]{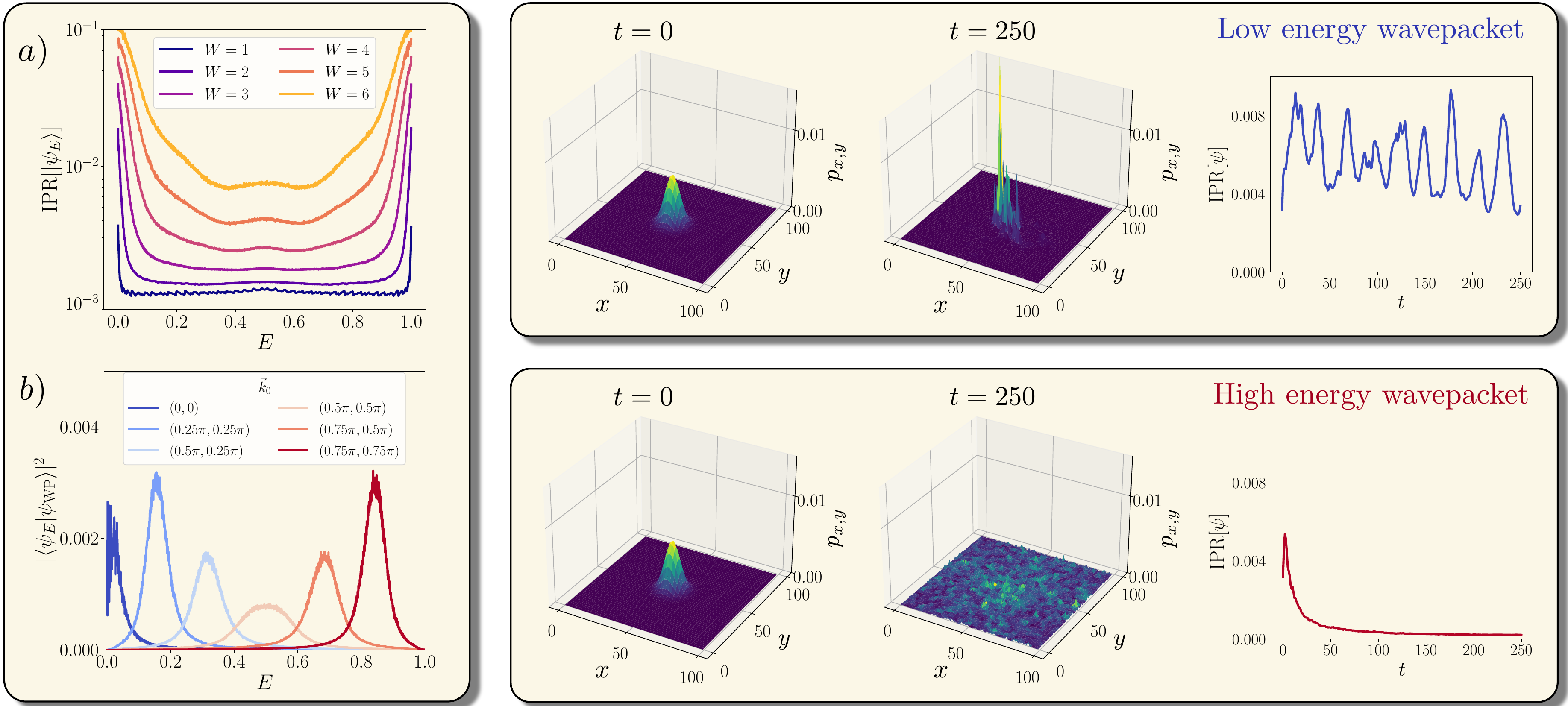}
    \caption{{\it Energy-dependent localization revealed through wavepacket dynamics.} Left panel: a) The IPR of the energy eigenstates $|\psi_E\rangle$ for a variety of disorder strengths $W$.
    The IPR is defined in Eq.~\eqref{eq:IPR}, and a larger value corresponds to a more localized wavefunction. 
    b) The overlap of wavepackets with a variety of momentum $\vec{k}_0$ onto energy eigenstates with $W=3$.
    The wavepacket is defined in Eq.~\eqref{eq:2dWP}, and a momentum spread of $\vec{\sigma}_p=(0.1,0.1)$ has been used.
    All results in this panel were computed for a $50\times 50$ lattice and averaged over 2000 disorder realizations.
    Energies have been rescaled to lie in the interval $E\in[0,1]$.
    Right panels: The time-evolution of the probability density $p_{x,y}$ of a low-energy wavepacket (top) with ${\vec k}_0 = (0,0)$ and a high-energy wavepacket (bottom) with ${\vec k}_0 = (0.75\pi,0.75\pi)$ at times $t=0$ and $t=250$ for a single disorder instance with $W=3$.
    The time dependence of the IPR (right plots) reveals that the low-energy wavepacket settles to a finite value, i.e. it remains localized, whereas the high-energy wavepacket quickly decays to a small value, i.e. it becomes delocalized.}
    \label{fig:Energy_dependent_WP}
\end{figure*}
A minimal 2D model that exhibits a finite-size mobility edge is the Anderson model~\cite{anderson1958absence} describing hard-core bosons hopping between sites with a random on-site potential. 
The Hamiltonian in second quantized form is,
\begin{align}
H \  &= \   -\sum_{\langle i,j \rangle} (b^{\dagger}_ib_j + b_i b_j^{\dagger}) \ +\ \sum_iW_i b^{\dagger}_ib_i\nonumber \\
&\to \  -\sum_{\langle i,j \rangle} (\sigma^+_i\sigma^-_j + \sigma^+_j\sigma^-_i) \ +\ \frac{1}{2}\sum_iW_i(1-Z_i) \ ,
\label{eq:H_tightbind_spin}
\end{align}
where $b_i$ and $b^{\dagger}_i$ are hard-core boson creation and annihilation operators.
These are mapped to spin operators in the second line such that (un)occupied sites correspond to $(|0\rangle)$ $|1\rangle$ in the computational z-basis.
The sum $\langle i,j\rangle$ runs over nearest neighbors on a $L_x \times L_y$ rectangular lattice with periodic boundary conditions (PBCs) and the sum on $i$ is over all sites. 
The disorder $W_i$ is drawn from a uniform random distribution $W_i \in [-W/2, W/2]$.
This Hamiltonian conserves the total particle number $N_e = \sum_i (1-Z_i)/2$, and this section will focus on the $N_e=1$, single particle, subspace.
The dimension of the single particle Hilbert space grows linearly with the system size, allowing classical simulations to be performed for large lattices.

For no disorder $W=0$, the system is translationally invariant and the energy eigenstates are single-particle plane waves,
\begin{align}
|\psi_{{\vec k}}\rangle \ = \ \frac{1}{L_x L_y}\sum_{x}\sum_{y}e^{i \vec{k}\cdot {\vec x}}|e_{x,y}\rangle \ ,
\label{eq:freePlaneWave}
\end{align}
with energies,
\begin{align}
E({\vec k}) \ = \ -2(\cos{k_x} + \cos{k_y}) \ .
\label{eq:freeDispersion}
\end{align}
The state $|e_{x,y}\rangle$ has a single particle located at site ${\vec x} = (x,y)$. 
As a bit string in the computational basis it has a single `1' on site $(x,y)$ with all else `0'.
The momentum lies in the first Brillouin zone, ${\vec k} = (k_x,k_y)$ with $k_{x,y} = 2 \pi n /L_{x,y}$ and $n$ an integer such that $k_{x,y} \in (-\pi,\pi]$.
Wavepackets are Gaussian superpositions of these momentum plane waves given by,
\begin{align}
|\psi_{\text{WP}}\rangle \ &= \ {\cal N}\sum_{\vec k} e^{- i {\vec k}\cdot {\vec x_0}}e^{-\frac{(k_x- k_{0,x})^2}{4\sigma_{p,x}^2}}e^{-\frac{(k_y- k_{0,y})^2}{4\sigma_{p,y}^2}}|\psi_{\vec k}\rangle \nonumber \\ 
&\equiv \ \sum_{x}\sum_{y} c_{x,y} |e_{x,y}\rangle \ .
\label{eq:2dWP}
\end{align}
The momentum is centered around $\vec{k}_0$ with spread $\vec{\sigma}_p$ and ${\cal N}$ is for normalization.
The wavefunction in position-space is implicitly defined in the second line where the magnitude of the coefficients $c_{x,y}$ follows a Gaussian centered at ${\vec x}_0=(x,y)$ with widths set by $1/\vec{\sigma_p}$.
Due to Eq.~\eqref{eq:freeDispersion}, states with a small spread in momentum also have a small energy variance.
For wavepackets supported on a constant fraction of the spatial volume the energy variance scales as $(\Delta E)^2 ={\mathcal O}(1/N)$ as shown in Appendix~\ref{a:deltaE}.

Disorder breaks translational invariance and the energy eigenstates are no longer plane waves.
Intuitively, the lowest-energy states tend to localize near sites with the most negative disorder while the highest-energy states tend to localize near sites with most positive disorder.
On a finite-size lattice, states away from the edges of the energy spectrum can be effectively delocalized, with wavefunctions spread across the whole lattice.
The localization of a wavefunction $|\psi\rangle$ can be quantified by the inverse participation ratio (IPR),
\begin{align}
\text{IPR}[\psi] \ = \ \sum_{x}\sum_{y} |\langle e_{x,y}|\psi\rangle|^4 \ .
\label{eq:IPR}
\end{align}
The IPR takes on values between $1/N$ and 1, corresponding to maximally delocalized (e.g. plane wave) and localized (e.g. a delta function) wavefunctions, respectively.
The IPR of the single-particle eigenstates in the 2D Anderson model for a range of disorder strengths is shown in Fig.~\ref{fig:Energy_dependent_WP}a).
The states near the edges of the spectrum are the most localized, while those in the middle of the spectrum are more delocalized.
This is a consequence of the mobility edge present in the finite-size 2D Anderson model.
Increasing the strength of the disorder leads to stronger localization as seen by the larger IPR across the spectrum.

Without disorder, the energy variance of a wavepacket can be made arbitrarily small by decreasing $\vec{\sigma_p}$.
Disorder increases the energy variance because the momentum plane waves that make up the wavepacket are no longer eigenstates.
This adds a term to the energy variance that scales as $(\Delta E)^2 ={\mathcal O}(NW^2)$ when averaged over disorder instances as shown in Appendix~\ref{a:deltaE}.
Therefore, wavepackets provide the best energy resolution for weak disorder.
This is illustrated in Fig.~\ref{fig:Energy_dependent_WP}b) where the overlap of wavepackets with several momenta ${\vec k}_0$ onto the energy eigenstates with disorder is shown.
Increasing the momentum shifts the support of the wavefunction onto eigenstates with higher energies.
Thus, even in the presence of disorder, increasing the wavepacket momentum increases its energy.

The existence of a mobility edge can be determined from the dynamics of wavepackets with different energies.
This is illustrated in the right panels of Fig.~\ref{fig:Energy_dependent_WP}, which show the time evolution of the probability density $p_{x,y}$ for both a low- and high-energy wavepacket.
The low-energy wavepacket has $\vec{k}_0=(0,0)$ and is primarily composed of localized eigenstates.
Therefore, its probability density does not spread significantly under time evolution.
In contrast, the high-energy wavepacket with $\vec{k}_0=(0.75\pi,0.75\pi)$ quickly delocalizes under time evolution.
The IPR as a function of time provides a robust signature of this energy-dependent localization and is also shown in the right panels of Fig.~\ref{fig:Energy_dependent_WP}.
The IPR of the low-energy wavepacket settles to a finite value that is similar to its value at $t=0$, whereas the high-energy wavepacket decays close to the minimum value of $1/N$.

\section{Error mitigation with Max-Likelihood Estimation}
\label{s:MLE}
\noindent
The particle number $N_e=1$ is conserved under dynamics generated by the Hamiltonian in Eq.~\eqref{eq:H_tightbind_spin}.
On a noisy quantum computer, there are errors that lead to measurements with $N_e\neq 1$.
A standard approach to mitigating these errors is to post-select on measured bit strings with $N_e=1$.
This can be an effective error mitigation strategy, see e.g. Ref.~\cite{Chernyshev:2025lil}, but comes with a shot overhead that scales inversely with the probability of no errors, which, for a constant error-rate per gate $p$ is roughly $(1-p)^{\# \text{gates}}$.
For dense quantum circuits, the shot overhead therefore increases exponentially with both system size and circuit depth (equivalently, simulation time), and post-selection quickly becomes impractical.
To overcome this limitation, we develop an error mitigation strategy that uses all of the shots and significantly outperforms post-selection.
Related strategies for mitigating bit-flip errors have been developed in several contexts.
These include correcting Gauss’s-law violations using minimum-weight matching~\cite{Cobos:2025krn}, correcting particle-number–violating bitstrings via probabilistic bit-flips in a self-consistent loop~\cite{Robledo-Moreno:2024pzz}, and applying MLE to constrained output distributions~\cite{Baron:2024kdb,Chandramouli:2025rfx}.

In the absence of noise, the outputs of our simulations are bit strings $e_i$ with a single `1' on qubit $i$. 
The underlying noiseless probability distribution is described by an $N$-dimensional vector ${\vec p}$.
The goal is to estimate ${\vec p}$ from a collection of measured noisy bit strings $\{b_j\}$ and their associated counts $\{c_j\}$, with $N_{\text{shots}} = \sum_j c_j$.
Because all of our measurements are in the $z$-basis, it suffices to only correct bit-flip errors acting on the final state.
Consider a quantum channel of IID bit-flip errors with strength $\epsilon$ acting on the final state density matrix $\rho$,
\begin{align}
&{ \cal E}(\rho) \ = \ \otimes_{i=0}^{N-1}\left [(1-\epsilon)\rho + \epsilon X_i\rho X_i \right ] \ .
\label{eq:Esimple}
\end{align}
A justification for this error model will be provided below.
The likelihood for observing the measured $\{b_j\}$ and $\{c_j\}$ from an underlying distribution ${\vec p}$ is,
\begin{align}
    &L(\vec{p},\epsilon,
    \{b_j\},\{c_j\}) \ = \nonumber \\
    &\ \sum_j \left (\sum_i p_i\epsilon^{d_H(b_j,e_i)}(1-\epsilon)^{N-d_H(b_j,e_i)}\right )^{c_j} \ ,
\end{align}
where $d_H(b_j, e_i)$ is the Hamming distance between the measured bit string $b_j$ and $e_i$.
This is the minimal number of bit-flips needed to map $e_i$ to $b_j$.
The estimated noise-free probability distribution and error rate is obtained by maximizing the likelihood function over ${\vec p}, \epsilon$,
\begin{align}
\hat{{\vec p}},\hat{\epsilon} \ = \ \text{argmax}\left [L(\vec{p},\epsilon,
    \{b_j\},\{c_j\}) \right ] \ . 
\end{align}
In practice, maximizing the log of the likelihood function improves numerical stability.

The success of our MLE-based error mitigation hinges on the accuracy of the uncorrelated bit-flip error model.
A more general error model would include, for example, a Pauli noise channel acting after each circuit layer.\footnote{Many error mitigation strategies currently assume a sparse Pauli noise channel acting after a layer of gates~\cite{Haghshenas:2025euj,Aharonov:2025ssq,Temme:2016vkz,Berg:2022ugn,Wack:2021gvg,Fischer:2024ipl}.}
However, even if Pauli errors are initially uncorrelated, propagation through subsequent circuit layers can build correlations.
Empirically, we find that these correlations can be ignored in our experiments.
This is likely because disorder also localizes errors, preventing them from spreading and becoming correlated when they have overlapping support.
For example, in the localized phase, it has been shown that operator spreading is bounded by a light-cone that grows only logarithmically in time~\cite{Huang_2016,chen2016universallogarithmicscramblingbody,Deng:2017pen}.
In a quantum simulation of Trotterized time evolution this limits the spreading of errors to be logarithmic with circuit depth.
Note that because of the limited error propagation, it is expected that dynamics in a localized phase will also be easier to classically simulate with Pauli path methods~\cite{Begusic:2023jwi,Broers:2024eoa,Rudolph:2025gyq}.

\section{Quantum Simulations on {\tt H2-2}}\label{s:simulations}
\begin{figure*}
    \centering
    \includegraphics[width=0.75\linewidth]{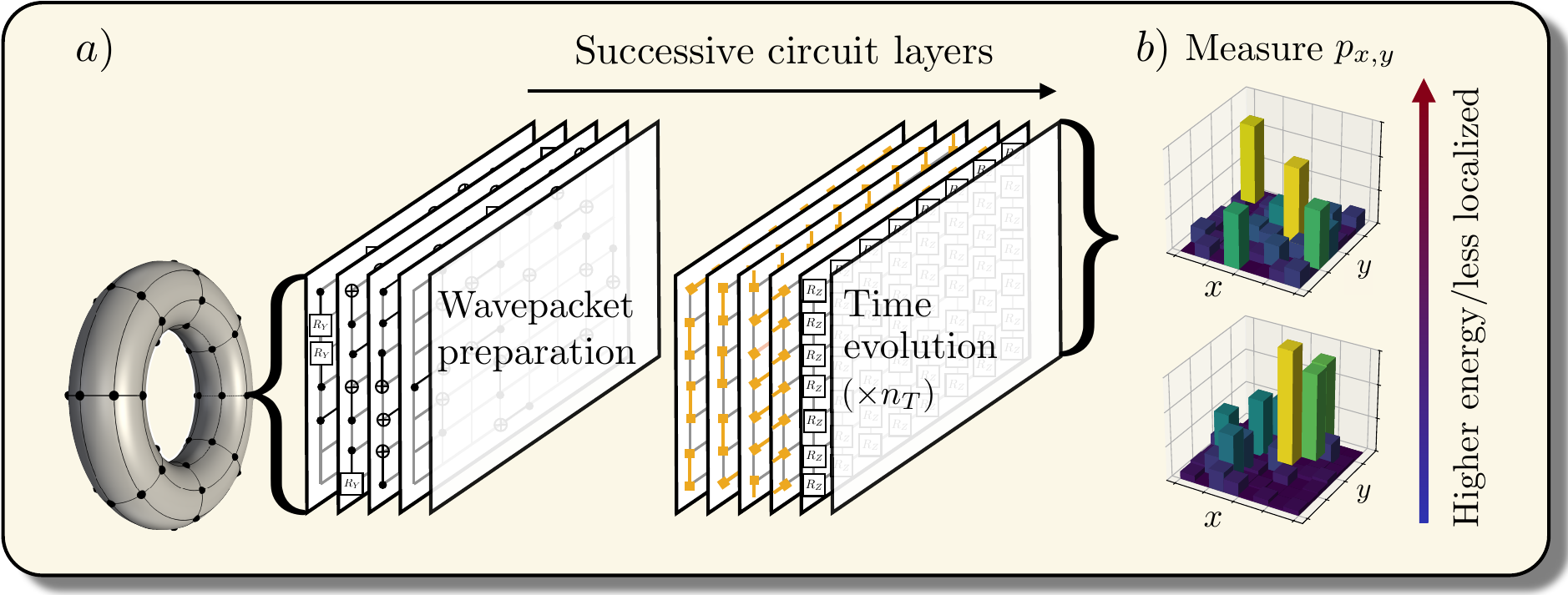}
    \caption{\textit{Overview of quantum simulations.}
    a) The 2D torus is mapped to 56 qubits on an $8\times7$ lattice. 
    The circuit begins with initialization layers that prepare the single-particle wavepacket of Eq.~\eqref{eq:2dWP} (see Appendix~\ref{a:qcircs} for their explicit construction). 
    Then, $n_T$ steps of Trotterized time evolution under the 2D Anderson Hamiltonian in Eq.~\eqref{eq:H_tightbind_spin} are applied.
    Each Trotter step has four layers of nearest-neighbor hopping terms  followed by a layer of $R_Z$ gates implementing the disorder.
    The hopping terms are represented by gold square dumbbells that implement $\exp(i \delta t (XX+YY)/2)$. 
    The decomposition into gates is given by the $XXZ(-\delta t,0)$ circuit block in Fig.~\ref{fig:singletWP}b).
    b) The single-particle probability density $p_{x,y}$ is determined from z-basis measurements.
    Low- (high-) energy wavepackets exhibit more (less) localized probability densities due to the finite-size mobility edge.}
    \label{fig:Quantum_results}
\end{figure*}
\noindent
In this section, we observe energy
dependent localization in the 2D Anderson model by simulating the dynamics of low- and high-energy wavepackets on Quantinuum's {\tt H2-2} quantum computer.
The simulations are performed on a $8\times7$ lattice with PBCs, fully utilizing all 56 qubits available on  {\tt H2-2}.
An overview of our quantum simulation strategy is illustrated in Fig.~\ref{fig:Quantum_results}.
First, a quantum circuit is applied that initializes the wavepacket.
This state is a superposition over all computational basis states with a single `1', with amplitudes specified by the wavepacket profile in Eq.~\eqref{eq:2dWP}.
The circuits for preparing this state build on Refs.~\cite{Cruz_2019,Farrell:2025nkx} and are given in Appendix~\ref{a:qcircs}.
The state preparation circuits require $2N$ two-qubit gates and a two-qubit gate depth of $2\lceil\log_2(N)\rceil$.\footnote{Mid-circuit measurements and feedforward (MCM-FF) can reduce the circuit depth required to prepare wavepackets at the cost of additional sample overhead~\cite{Farrell:2025nkx}.
A comparison of MCM-FF and unitary wavepacket preparation methods using Quantinuum's {\tt H2-2E} noisy classical emulator is given in Appendix~\ref{a:qcircs}. 
It is found that the unitary method currently performs the best.}

Once prepared, the wavepacket is evolved under the Hamiltonian in Eq.~\eqref{eq:H_tightbind_spin} using a first-order Trotterization of $e^{- i \hat{H} t}$.
The ordering of the terms is shown in Fig.~\ref{fig:Quantum_results}a).
The hopping terms are split into four commuting sets and implemented sequentially.
The disorder is implemented at the end of each Trotter step with single-qubit $R_Z$ rotations.
A Trotter step size of $\delta t=0.25$ is chosen to balance Trotter errors against the number of Trotter steps $n_T$ required to reach a simulation time $t$.
After the time evolution the qubits are measured in the computational basis to determine the probability density $p_{x,y}$ as shown in Fig.~\ref{fig:Quantum_results}b).
The IPR is then computed from the probability density using Eq.~\eqref{eq:IPR}.

A disorder strength of $W=6$ is chosen to be large enough for the spectrum to exhibit energy-dependent localization, but small enough for the wavepackets to have minimal energy variance.
The simulations are all performed using a single instance of the random disorder strengths that captures the qualitative features of the disorder-averaged results.
The low and high energy wavepackets have momentum $\vec{k}_0 = (0,0)$ and $\vec{k}_0 = (0.5\pi,-0.1\pi)$ respectively, which were found to give the strongest signals of energy-dependent localization.\footnote{
Components of the initial wavepacket with probabilities $p_{x,y}<0.01$ were truncated for simulation times $t=0,1$ giving initial states supported on 36 and 32 qubits for the low- and high-energy wavepackets, respectively.
The effects of this truncation are negligible, and the high-energy wavepacket simulation at $t=0$ was used to benchmark {\tt H2-2} against the classical {\tt H2-2E} emulator that is limited to 32 qubits.}
Further details on the choice of quantum simulation parameters are provided in Appendix~\ref{a:qsim_params}.

\begin{table}[t]
\centering
\renewcommand{\arraystretch}{1.4}
\begin{tabularx}{\linewidth}{|c||Y|Y|Y|Y|Y|} \hline
\makecell{$t$}  & \makecell{\# of two-\\qubit gates} & \makecell{Low-energy \\ PS \%} & \makecell{High-energy \\ PS \%}  & \makecell{\# of shots} \\ \hline\hline
0 & 70/62 & 88.4 & 91.3 & 500/1000 \\ \hline
1 & 966/958 & 29.6 & 45.2 & 250 \\ \hline
2 & 1,902 & 20.0 & 19.0 & 400 \\ \hline
3 & 2,798 & -- & 14.7 & 450 \\ \hline
\end{tabularx}
\renewcommand{\arraystretch}{1}
\caption{For a simulation time $t$, column 2 gives the total number of two-qubit gates, columns 3 and 4 give the percentage of shots that survive post-selection (PS) on measurements with $N_e=1$ for the low- and high-energy wavepackets, respectively.
The total number of shots is given in column 5. 
Quantities separated by a `$/$' correspond to low and high energy wavepackets, respectively.}
\label{tab:PS_rates}
\end{table}

\begin{figure}
    \centering
    \includegraphics[width=\linewidth]{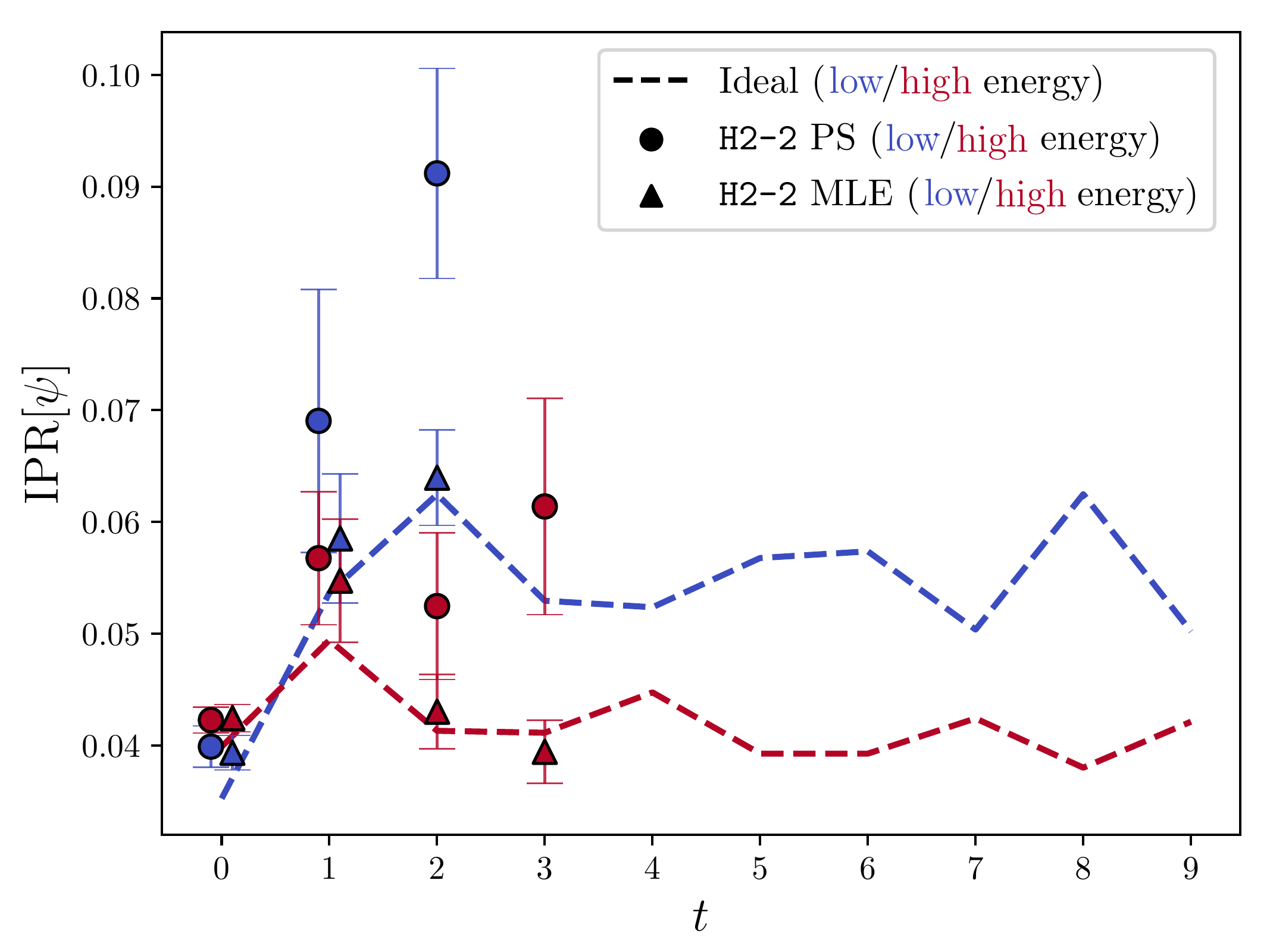}
    \caption{{\it The IPR obtained from {\tt H2-2}.}
    The IPR evaluated at a selection of simulation times for both the low-energy (blue) and high-energy (red) wavepackets.
    The IPR is defined in Eq.~\eqref{eq:IPR}, and a larger value corresponds to a more localized wavefunction.
    The dashed lines are the results of noiseless classical simulations. 
    The points with error bars are the results obtained from Quantinuum's {\tt H2-2} quantum computer, and at $t=0,1$ have been offset for clarity.
    The circles and triangles correspond to error mitigation using post-selection (PS) and max-likelihood estimation (MLE), respectively.
    The error bars are statistical and are determined by bootstrap resampling over the measured bit strings. These results are tabulated in Table~\ref{tab:quantum_result_IPR}.}
    \label{fig:IPR_results}
\end{figure}

The time evolution of the IPR of the low and high energy wavepackets obtained from {\tt H2-2} is shown in Fig.~\ref{fig:IPR_results}.
The results of noiseless classical simulations (dashed lines) are compared to the {\tt H2-2} results post-processed with post-selection (PS) (circles) and MLE (triangles) error mitigation.
With PS, measurements that do not conserve the particle number of the initial state, $N_e =1$, are thrown out.
For each simulation time, the total number of shots and the post-selection rates are given in Table~\ref{tab:PS_rates}.
Due to the increasing number of two-qubit gates, the post-selection rate decreases with simulation time, reaching a minimum of $14.7\%$ at $t=3$.
Only the high-energy wavepacket was simulated at $t=3$ due to limited allocated runtime on {\tt H2-2}.
The MLE method is described in Section~\ref{s:MLE} and uses all of the shots to predict the noiseless output probability distribution.

The noiseless classical simulations show that the low-energy wavepacket has a larger IPR than the high-energy wavepacket for $t\geq 1$, consistent with the presence of a finite-size mobility edge.
The PS results also follow this trend, but with large error bars and an IPR that is systematically larger than the noiseless result.
This systematic error is a consequence of undersampling the wavefunction due to discarding shots during post-selection.
Undersampling inhibits the resolution of small wavefunction amplitudes and leads to predictions that are biased toward a higher IPR.
Error mitigation with MLE provides a substantial improvement. 
The error bars become smaller, and deviations from the expected result are consistent with statistical fluctuations.
This highlights that wavepacket-based probes of energy-dependent localization are especially effective when coupled with MLE error mitigation.
The probability density predicted by MLE is also consistent with noiseless simulations, as shown in Appendix~\ref{a:qsim_details}.
%

\section{Preparing wavepackets in the 1D \texorpdfstring{$XXZ$}{} model}\label{s:halffilling}
\noindent 
Having demonstrated the ability of single-particle wavepackets to identify energy-dependent localization transitions, we now address a key challenge in extending these methods to interacting many-body systems: the preparation of wavepackets.
In this section, we present a quantum algorithm for preparing wavepackets in a one-dimensional model of interacting fermions.

Consider a one-dimensional system of $N_e=N/2$ spinless fermions, i.e at half filling, with nearest-neighbor density-density interactions and PBCs.
The Hamiltonian is,
\begin{align}
    &H \ = \  \sum_{i=0}^{N-1}  c_i^{\dagger}c_{i+1}+c_{i+1}^{\dagger}c_i + 2\Delta c_i^{\dagger}c_i c_{i+1}^{\dagger}c_{i+1}\nonumber  \\ 
    &\to -\frac{1}{2}\sum_{i=0}^{N-1} \left ( X_iX_{i+1}+Y_iY_{i+1} -\Delta Z_i Z_{i+1}\right )  \ , 
    \label{eq:hamXXZ}
\end{align}
where $c_i^{\dagger}$ and $c_i$ are fermionic creation and annihilation operators respectively and $\Delta$ is the interaction strength. 
In the second line, the Jordan-Wigner transformation~\cite{jordan:1928wi} has been used to map the fermionic operators onto spins.\footnote{Global transformations have been performed on the fermionic operators to put the Hamiltonian into $XXZ$ form.
Also, terms have been dropped that are proportional to $\sum_i Z_i=N_e$ as these are a constant at half-filling~\cite{10.1093/acprof:oso/9780198525004.001.0001}. Lastly, the system size is assumed to be $N=2+4n$ so the Jordan-Wigner-mapped spin chain has PBCs.}
For $\Delta \leq 1$, the system is gapless, and its low-energy behavior in the thermodynamic limit is described by a Luttinger liquid with collective density excitations that behave as bosonic quasiparticles~\cite{10.1093/acprof:oso/9780198525004.001.0001}.
This will be the regime considered in the rest of this section.

\begin{figure}[htbp]
    \centering
    \includegraphics[width=\linewidth]{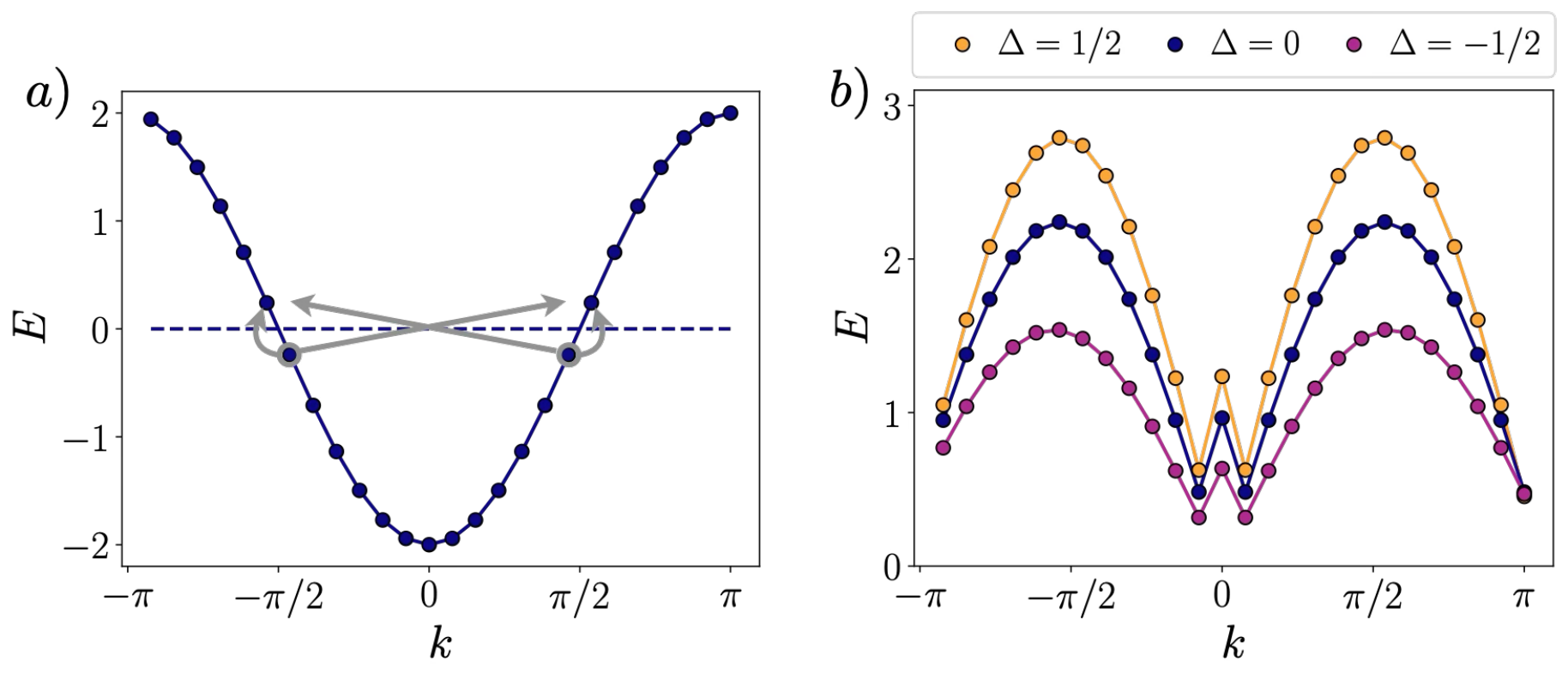}
    \caption{a) The single-particle energy spectrum in the 1D $XXZ$ model with no interactions.
    The ground state at half-filling has all states below the Fermi surface (dashed line) occupied.
    There are four degenerate lowest energy excitations corresponding to the particle hole excitations represented by the gray arrows.
    b) The many-body spectrum of the lowest energy excitations for each momentum (the $|\psi_k\rangle$).
    The energies are determined from exact diagonalization for $N=26$ and are shown for a selection of interaction strengths $\Delta$.
    The ground state energy has been subtracted.}
    \label{fig:EnergySpectrum}
\end{figure}
Wavepackets are defined analogously to Eq.~\eqref{eq:2dWP},
\begin{align}
|\psi_{\text{WP}}\rangle \ &= \ {\cal N}\sum_{k} e^{- i k x_0}e^{-(k-k_0)^2/(4\sigma_p^2)} |\psi_{k}\rangle \ ,
\label{eq:1DWP}
\end{align}
where $|\psi_{k}\rangle$ is the lowest-energy eigenstate with momentum $k$.
This definition of $|\psi_{k}\rangle$ strictly breaks down in the thermodynamic limit, where there is no energy gap separating $|\psi_{k}\rangle$ from the continuum of states at the same momentum.
However, for finite $N$, there is an energy gap, and the $|\psi_k\rangle$ can be thought of as single quasiparticle excitations of the many-body system.\footnote{
There are degeneracies in the spectrum that inhibit the unique identification of $|\psi_k\rangle$ for $\Delta=0$ and $\Delta=-1/2$.
This is due to an enhanced $sl_2$ loop algebra symmetry~\cite{Deguchi:1999rq} that causes a two-fold degeneracy at each momentum for $|k|>2\pi/N$ ($\Delta = 0$) and $|k|>4\pi/N$ ($\Delta = -1/2$).}

The goal is to determine circuits that approximately prepare $|\psi_{\text{WP}}\rangle$ on a quantum computer.
Some features of the $|\psi_{k}\rangle$ that will be important for preparing $|\psi_{\text{WP}}\rangle$ are present without interactions.
In the non-interacting limit, the many-body energy eigenstates are Slater determinants of $N_e = N/2$ single-particle plane-wave eigenstates.
Each eigenstate is specified by a set of occupied momenta $\{k_i\}$ and its energy is a sum over the single-particle energies,
\begin{align}
E \ = \ -2\sum_{\{k_i\}}\cos(k_i)  \ . 
\end{align}
The ground state has the $N_e = N/2$ lowest energy eigenstates filled corresponding to all of the states below the Fermi surface, see Fig.~\ref{fig:EnergySpectrum}a).
We will assume $N=2+4n$ in this section so that the momentum of the occupied single-particle states sums to zero. 
This choice ensures that the total momentum of the ground state is $k=0$.
The lowest energy excitation corresponds to a particle-hole excitation where the hole is right below the Fermi surface and the particle is right above the Fermi surface. 
There are four such degenerate excitations with energy $E=4\sin(\pi/N)$ and momentum $k=[\pm\frac{2\pi}{N}, \pm\pi]$.
These particle-hole excitations are illustrated by the gray arrows in Fig.~\ref{fig:EnergySpectrum}a). 

Turning on interactions changes the spectrum of the many-body excitations, which are no longer simple Slater determinants of single-particle eigenstates.
At half-filling, the $|\psi_k\rangle$ are defined to be the lowest-energy excitation for each momentum.
The energies of these states is shown in Fig.~\ref{fig:EnergySpectrum}b) for several values of the interaction strength.
In general, the excitations $|\psi_k\rangle$ and $|\psi_{-k}\rangle$ are degenerate due to time reversal symmetry.
In the non-interacting case $\Delta = 0$, there is a larger set of degeneracies relating the $|\psi_k\rangle$ with momentum,
\begin{align}
k \ = \ \pi\left (\frac{1}{2} + \frac{1}{N}\right )\pm \frac{2\pi n}{N} \   , \   n\in \left \{1,2,...\frac{1}{2}\left (\frac{N}{2}-1 \right ) \right \} \ , 
\end{align}
as well as those with $k\to-k$.
These degeneracies correspond to reflections about the maxima of the dispersion in Fig.~\ref{fig:EnergySpectrum}b), together with $E(k) = E(-k)$.
For $\Delta \neq 0$, the degeneracy relating energies related by reflection is lifted.
Nevertheless, the hierarchy
\begin{align}
E(k) < E(k\pm \pi)\quad \text{for} \quad |k| <\pi/2
\label{eq:energy_observation}
\end{align}
is preserved.
This observation is crucial, as it allows the preparation of wavepackets supported on $|k|<\pi/2$ to be reformulated as an energy minimization problem.

Our method for preparing the wavepacket in Eq.~\eqref{eq:1DWP} on a quantum computer builds off the ground state preparation algorithm given in Ref.~\cite{Yu:2022ivm}, which we briefly review.
The Hamiltonian in Eq.~\eqref{eq:hamXXZ} can be split into the terms that couple even-odd numbered sites and odd-even numbered sites.
The ground state of the terms that couple even-odd sites is a tensor product of triplet states, $|\text{triplet}\rangle^{\otimes N/2}$,
where
\begin{align}
&|\text{triplet}\rangle \ = \ \frac{1}{\sqrt{2}}\left (|01\rangle +|10\rangle \right ) \ .
\end{align}
Excited states can be accessed by replacing a triplet with a singlet,
\begin{align}
&|\text{singlet}\rangle \ = \ \frac{1}{\sqrt{2}}\left (|01\rangle -|10\rangle \right ) \ . 
\end{align}
Starting from $|\psi_{\text{init}}\rangle = |\text{triplet}\rangle^{\otimes N/2}$, the ground state of the total Hamiltonian can be prepared by adiabatically turning on the terms in the Hamiltonian that couple odd-even sites.
To reduce the circuit depth, the time steps in the adiabatic preparation circuit are replaced by variational parameters $\vec{\theta}$.
This defines a parameterized quantum circuit,
\begin{align}
U(\vec{\theta}) = &\prod_{l =0}^{n_L-1}\bigg ( \prod_{n \ \text{odd}} e^{-\frac{i}{2}\left [ \theta_{4l+2}(X_nX_{n+1}+Y_nY_{n+1})+\theta_{4l+3} Z_nZ_{n+1}\right ]} \nonumber \\
& \prod_{n \ \text{even}}e^{-\frac{i}{2}\left [ \theta_{4l}(X_nX_{n+1}+Y_nY_{n+1})+\theta_{4l+1} Z_nZ_{n+1}\right ]} \bigg ) \ ,
\label{eq:ansatz_circ}
\end{align}
where $n_L$ is the number of circuit layers in the ansatz circuit.
The $\vec{\theta}$ are optimized to minimize the energy
\begin{align}
E_{\text{ansatz}} \ = \ \langle \psi_{\text{init}}|U^{\dagger}(\vec{\theta})\, H\, U(\vec{\theta})|\psi_{\text{init}}\rangle\ ,
\label{eq:Eansatz}
\end{align}
and the energy minimum is a good approximation to the ground state.

This method can be extended to the preparation of wavepackets using ideas from Ref.~\cite{Farrell:2025nkx}.
To begin, consider the preparation of $|\psi_{k}\rangle$.
In principle, this can be done by initializing some state with  momentum $k$ and $N_e=N/2$, and then variationally minimizing the energy using a circuit ansatz that is translationally invariant and conserves $N_e$.\footnote{
A translationally invariant circuit can be constructed from a product of unitaries that are generated by translationally hermitian operators, e.g. $U = \prod_k e^{i \theta_k O_k}$ where $O_k = \sum_i O_{k,i}$.}
Translational invariance guarantees that the circuit does not change the momentum of the initial state, i.e. that the prepared state always has momentum $k$.
The challenge is to find a circuit ansatz that prepares $|\psi_{k}\rangle$ with minimal quantum resources and has a favorable optimization landscape free of, e.g., local minima and barren plateaus~\cite{Wang:2020yjh}.
In Ref.~\cite{Farrell:2025nkx}, the circuit ansatz was constructed using ADAPT-VQE~\cite{Grimsley:2018wnd} equipped with a translationally invariant operator pool~\cite{Farrell:2023fgd,VanDyke:2022ffj} generated from the Hamiltonian Lie algebra.
In the gapless $XXZ$ model, we find that the optimization of this type of circuit ansatz gets stuck in local minima and does not converge to $|\psi_k\rangle$.
This may be due to the presence of long-range correlations in the target wavefunction, a consequence of the system being gapless, which cannot be easily constructed by a greedy algorithm like ADAPT-VQE.
One potential way to overcome this challenge is to include mid-circuit measurements and feedforward into the variational ansatz.
This has been shown to improve the loss landscape~\cite{Puente:2024bxt,Deshpande:2024kpt}, as well as the ansatz's ability to generate long range correlations~\cite{Alam:2024mit,Yan:2024xev,Niu:2024oxx}.
An exploration of this possibility is left for future work.

Instead, we use the $U(\vec{\theta})$ ansatz in Eq.~\eqref{eq:ansatz_circ}.
We suspect that this ansatz structure is more effective than ADAPT-VQE because it is related to an adiabatic path that can prepare the long-range correlations in the ground state.
This circuit preserves $N_e$ but is only invariant under translations by two lattice sites.
It therefore couples momentum modes $k$ and $k\pm\pi$.
Fortunately, this is not a problem as long as $|k|<\pi/2$; due to Eq.~\eqref{eq:energy_observation} the energy minimum of Eq.~\eqref{eq:Eansatz} will still be $|\psi_k\rangle$.

Our choice of initial state is motivated by the observation that
low-energy excitations can often be generated by acting with plane waves of local operators on top of the ground state~\cite{PhysRev.94.262,Haegeman:2013xcv}.
Therefore, we choose an initial state that is related by local perturbations to $|\text{triplet}\rangle^{\otimes N/2}$, the state that can be adiabatically connected to the ground state by $U(\vec{\theta})$.
Consider a plane wave of singlet excitations on top of a triplet background,
\begin{align}
&|k\rangle  =  \nonumber \\
&\sqrt{\frac{2}{N}}\sum_{n=0}^{N/2 -1} e^{2ik n}|\text{triplet}\rangle^{\otimes N/2-n-1}  |\text{singlet}\rangle  |\text{triplet}\rangle^{\otimes n} \ . 
\end{align}
We choose $|\psi_{\text{init}}\rangle = |k\rangle$ as the state that is superposition of the singlet excitation translated every two sites as this is easy to prepare on a quantum computer (discussed below).
As a consequence, $|\psi_{\text{init}}\rangle$ contains both momentum $k$ and $k\pm\pi$ components.

Next, the parameters in $U(\vec{\theta})$ are optimized to minimize the energy in Eq.~\eqref{eq:Eansatz}.
The circuit couples the $k$ and $k+ \pi$ components of $|\psi_{\text{init}}\rangle$ and,
due to Eq.~\eqref{eq:energy_observation}, the energy minimum is $|\psi_k\rangle$ as long as $|k|<\pi/2$.
In practice, as the energy decreases, more of the wavefunction amplitude is transferred from components with momentum $k\pm\pi$ to components with momentum $k$.

The generalization to wavepacket preparation is immediate.
Instead of a plane wave of spin singlet excitations initialize a wavepacket,
\begin{align}
&|\psi_{\text{init}}\rangle  =  {\cal N}\sum_{k} e^{- i k x_0}e^{-(k-k_0)^2/(4\sigma_p^2)} |k\rangle \nonumber \\
& \equiv\sum_{n=0}^{N/2 -1} c_{n}|\text{triplet}\rangle^{\otimes N/2-n-1}  |\text{singlet}\rangle  |\text{triplet}\rangle^{\otimes n} \ . 
\label{eq:singletWPinit}
\end{align}
The $c_n$ have a Gaussian profile in position space that specify the momentum content (must be supported on $0<|k|<\pi/2$).
The parameters in $U(\vec{\theta})$ are then optimized to minimize the energy in Eq.~\eqref{eq:Eansatz}, using as an initial guess an adiabatic turn on,
\begin{align}
&\theta_{4l} = 1/n_L\ , \quad &&\theta_{4l+1} = \Delta/n_L \ , \nonumber \\
&\theta_{4l+2} = \left ( l+1/2\right )/n_L^2 \ , \quad  &&\theta_{4l+3}= \Delta(l+1/2)/n_L^2 \ .
\end{align}
The translational symmetry of $U(\vec{\theta})$ implies that it acts independently on each $(k,k\pm\pi)$ momentum sector in a way that preserves the amplitude $(e^{-(k-k_0)^2/(4\sigma_p^2)})$ established by $|\psi_{\text{init}}\rangle$. 
Combined with the energy hierarchy in Eq.~\eqref{eq:energy_observation}, this implies that the variational ansatz is projecting each momentum component onto its lowest energy state, $|\psi_k\rangle$.
As a result,
the energy minimum in Eq.~\eqref{eq:Eansatz} is precisely the wavepacket in Eq.~\eqref{eq:1DWP}.\footnote{
The circuit does not satisfy $U(\vec{\theta}) = U(\vec{\theta})^*$ and can therefore add relative phases between the momentum modes in the wavepacket.
We find that this effectively translates the center of the wavepacket away from the specified $x_0$.}

\begin{figure}
    \centering
    \includegraphics[width=\linewidth]{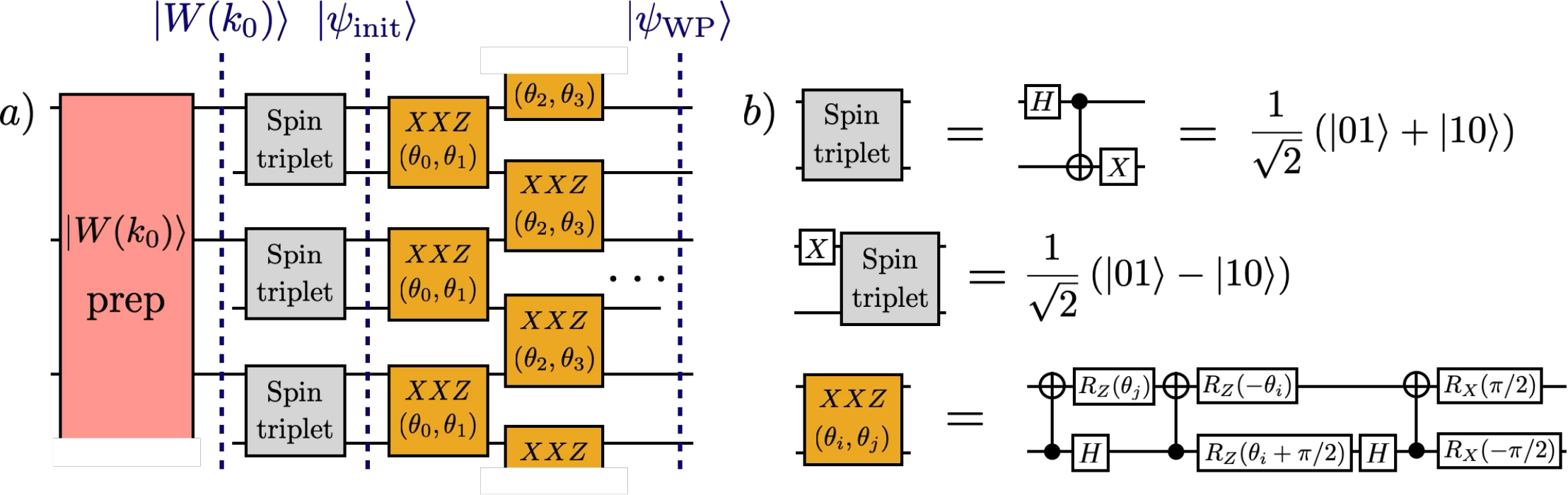}
    \caption{a) The structure of a quantum circuit that prepares a quasiparticle wavepacket in the 1D $XXZ$ model. 
    The wavefunction at intermediate points are marked with blue dashed lines and defined in the main text.
    b) The definition of the circuit blocks, with $XXZ(\theta_i, \theta_j) = e^{-\frac{i}{2}\left [ \theta_i(XX+YY)+\theta_j ZZ\right ]}$.
    For $\theta_i=-\delta t,\theta_j=0$ the gold circuit simplifies to two CNOTs and implements the evolution of the nearest neighbor hopping term represented by the gold square dumbbell in Fig.~\ref{fig:Quantum_results}.
    Circuits that prepare $|W(k_0)\rangle$ are given in Appendix~\ref{a:qcircs}.
    }
    \label{fig:singletWP}
\end{figure}
The first step of the wavepacket preparation algorithm is to apply a circuit that initializes $|\psi_{\text{init}}\rangle$.
An important observation is that the only difference between the circuit that prepares $|\text{singlet}\rangle$ and $|\text{triplet}\rangle$ is the insertion of an $X$ gate on the first qubit, see Fig.~\ref{fig:singletWP}b).
This implies that the wavepacket of singlet excitations in Eq.~\eqref{eq:singletWPinit} can be prepared by first acting with a superposition of $X$ gates on the even numbered qubits, and then applying the triplet preparation circuit in Fig.~\ref{fig:singletWP} across all even-odd numbered qubits.
The superposition of $X$ gates is equivalent to initializing
\begin{align}
|W(k_0)\rangle \ &= \ \sum_{n=0}^{N/2-1} c_{n}|e_{2n}\rangle \ ,
\end{align}
where the $c_n$ are given by Eq.~\eqref{eq:singletWPinit}.
This state has the same structure as the $N_e=1$ wavepackets in Section~\ref{s:preliminaries} and can be prepared in $2\lceil\log_2(N/2)\rceil$ two-qubit gate depth, see Appendix~\ref{a:qcircs} for details. 
The quantum circuits used to implement $U(\vec{\theta})$ and prepare $|\psi_{\text{WP}}\rangle$ are shown in Fig.~\ref{fig:singletWP}a).

\begin{figure}
    \centering
    \includegraphics[width=0.75\linewidth]{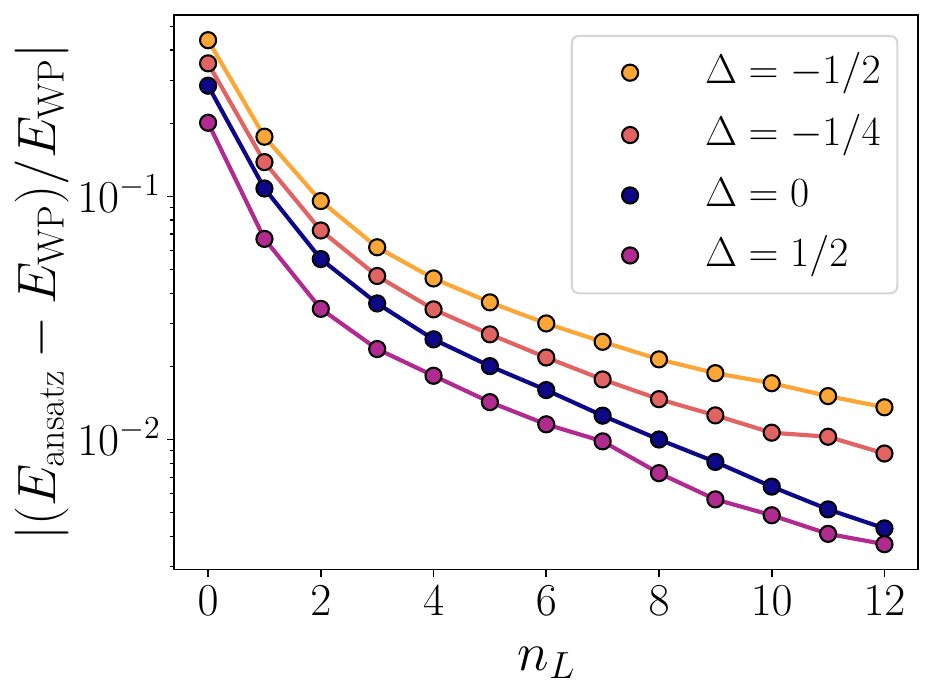}
    \caption{The difference between the energy of the prepared wavepacket $E_{\text{ansatz}}$ and exact wavepacket $E_{\text{WP}}$ (determined from exact diagonalization) for an increasing number of ansatz circuit layers $n_L$.
    Results for several interaction strengths $\Delta$ are shown.}
    \label{fig:1D_WP_Energies}
\end{figure}
We verify the wavepacket preparation procedure using a classical statevector simulator on $N=22$ sites.
The convergence of the ansatz energy to the exact wavepacket energy for an increasing number of variational layers $n_L$ is shown in Fig.~\ref{fig:1D_WP_Energies}.
We observe faster convergence for $\Delta>0$, likely because the initial state is closer in energy to the exact wavepacket.
Notably, although the ansatz energy approaches the exact value, the fidelity with the target state remains small.
This is expected in a gapless system because at each momentum there are many excitations close in energy to $|\psi_k\rangle$.
As a result, the prepared state should be understood as a wavepacket composed of low-energy excitations rather than the lowest-energy excitation at each momentum.

\begin{figure}
    \centering
    \includegraphics[width=0.75\linewidth]{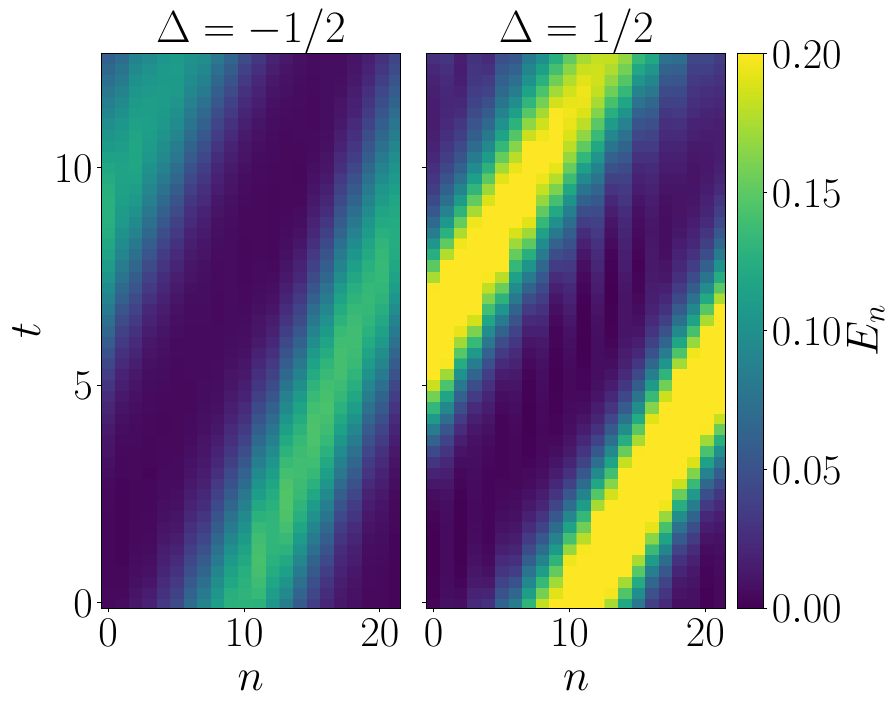}
    \caption{The dynamics of a quasiparticle wavepacket prepared from $n_L=12$ variational layers.
    The ground state subtracted energy density, $E_n$, is shown as a function of space and time with $\Delta = -1/2$ (left) and $\Delta = 1/2$ (right).
    Classical statevector simulations were used in both a) and b) and the simulation parameters are $k_0 = 0.25\pi$ and $\sigma_p = 0.2$, $N=22$.}
    \label{fig:1D_WP_Dynamics}
\end{figure}
Despite this, the prepared state still exhibits ballistic propagation.
This is confirmed through classical statevector simulations of time evolution, shown in Fig.~\ref{fig:1D_WP_Dynamics}b).
The ground-state-subtracted energy density $E_n$ for both $\Delta=-1/2$ and $\Delta=1/2$ is plotted throughout the dynamics. 
At $t=0$, the energy density follows a Gaussian profile, while subsequent time evolution produces propagation at an approximately constant velocity.
The propagation is faster for $\Delta=1/2$ than $\Delta=-1/2$ due to a larger group velocity $dE/dk$ in the regime of repulsive interactions.
This is consistent with the slope of the spectrum in Fig.~\ref{fig:EnergySpectrum}.
Together, these results demonstrate that this protocol can successfully prepare propagating wavepackets in a gapless many-body system.

\section{Outlook}\label{s:discussion}
\noindent
In this work, we have proposed using wavepacket dynamics as a tool for studying energy-resolved transport on quantum computers. 
This approach was demonstrated in the 2D Anderson model, where wavepacket dynamics reveal the presence of a finite-size mobility edge: 
low-energy wavepackets remain localized under time evolution, while high-energy wavepackets spread out. 
Experiments using Quantinuum’s {\tt H2-2} quantum computer confirm that this signature of a mobility edge can be robustly extracted from noisy quantum simulations. 
Looking toward quantum simulations in the many-particle regime, we introduced a quantum algorithm for preparing quasiparticle wavepackets in a one-dimensional model of interacting fermions.
Reframing wavepacket preparation as an energy-minimization problem enabled the use of relatively low-depth parameterized quantum circuits. 
Together, these results establish a pathway toward energy-resolved dynamical simulations in interacting many-body systems using a quantum computer.

Important open questions remain regarding the efficient preparation of wavepackets in many-body systems on quantum computers. 
Recent work~\cite{Maskara:2025oab,Constantinides:2025kjx} has shown that the fermionic Fourier transform can be efficiently implemented in systems with all-to-all connectivity and mid-circuit measurements.
This reduces the circuit depth required to prepare ground states of free fermion Hamiltonians from ${\mathcal O}(N)$~\cite{Kivlichan:2018bqq,Derby:2020aha} to ${\mathcal O}(\log N)$. 
More generally, it should be possible to prepare coherent superpositions of momentum-space occupations—such as wavepackets—in the free-fermion basis before Fourier transforming back to position space. Starting from such a free-fermion wavepacket could significantly reduce the depth of the energy-minimization circuit required to prepare an interacting wavepacket.

An alternative approach is to adiabatically turn on fermion–fermion interactions starting from a superposition of free-fermion momentum modes. 
This would enable the preparation of more general wavepackets that remain spatially localized but are composed of higher-energy excitations. 
Such capabilities would unlock the study of transport properties at energies spanning the full many-body spectrum, overcoming the present work’s limitation to energies within the single-quasiparticle band.
However, in disordered systems, the utility of wavepacket dynamics would remain limited, since maintaining a fixed fraction of the volume in the ground state necessarily leads to an energy variance that scales as $(\Delta E)^2 \sim \mathcal{O}(W^2 N)$ for large $N$, as shown in Appendix~\ref{a:deltaE}.
Computational basis states have the same ${\mathcal O}(N)$ scaling, implying that wavepacket-based approaches would provide an advantage primarily in the weak-disorder regime.
Note that this discussion has not considered transport studies where the energy resolution comes from controlling the trajectory of an external ``classical" field or particle, as has been done to study energy loss in lattice gauge theories~\cite{Farrell:2024mgu,Li:2025sgo,Barata:2025hgx}.

Another important future direction is to generalize the MLE error mitigation strategy employed in this work. 
In particular, extending the underlying error model beyond IID bit-flip errors would allow for a more realistic treatment of noise, 
including the estimation of systematic uncertainties arising from neglected error correlations beyond a given distance.
A further challenge concerns the scalability of the current MLE procedure. 
At present, the optimization is performed over all output bit strings consistent with symmetries. 
While this space grows only linearly with system size for single-particle states, it grows exponentially in the many-particle case, rendering the optimization intractable. 
One potential way to address this limitation would be to instead optimize over the space of measured bit strings, and then systematically enlarge the optimization domain by including bit strings that differ from those measured by one bit-flip, two bit-flips, etc.
Demonstrating convergence of the reconstructed observables under such an expansion would provide a controlled and scalable error-mitigation strategy for larger systems.

\vspace{2em}


\begin{acknowledgements}
\noindent
We would like to thank Jessica Jiang, Ruoshi Jiang, Bibek Pokharel, John Preskill, Ilan Rosen, Kevin Smith, Federica Surace and Nikita Zemlevskiy for valuable discussions.
M.L. is funded in part by Caltech’s Institute for Quantum Information and Matter (IQIM) WAVE fellowship program. 
R.F. acknowledges support from the U.S. Department of Energy QuantISED program through the theory consortium “Intersections of QIS and Theoretical Particle Physics” at Fermilab, from the U.S. Department of Energy, Office of Science, Accelerated Research in Quantum Computing, Quantum Utility through Advanced Computational Quantum Algorithms (QUACQ), and from the Institute for Quantum Information and Matter, an NSF Physics Frontiers Center
(PHY-2317110). 
R.F. additionally acknowledges support from a Burke Institute prize fellowship.
This research used resources of the Oak Ridge Leadership Computing Facility, which is a DOE Office of Science User Facility supported under Contract DE-AC05-00OR22725.
\end{acknowledgements}

\onecolumngrid
\begingroup
\hypersetup{linkcolor=black}
\endgroup

\newpage

\appendix
\section{Energy variance of wavepackets}
\label{a:deltaE}
\noindent
The energy variance of the initial wavepacket sets the energy resolution that can be achieved. 
In this appendix the energy variance of wavepackets is computed assuming that their size is a fixed fraction of the volume, i.e. scales with system size, and that the momentum support is away from the edges of the Brillouin zone, $k=\pm \pi$.
The focus is on determining how the energy variance scales in the large volume limit, $N\to \infty$.
The dispersion relation is left general until the end when specific results are provided for the single-particle case.

The Hamiltonian that we consider breaks into two pieces,
\begin{align}
H \ = \ H_0 \ + \ H_W
\end{align}
where $H_0$ is the part of the Hamiltonian containing the kinetic and particle-particle interactions terms, and $H_W$ contains the terms with random on-site disorder.
For simplicity and ease of notation we start with one dimension. 
The generalization to higher dimensions is straightforward and is given at the end.
The goal is to determine the energy variance averaged over disorder instances, defined as,
\begin{align}
(\Delta E)^2 \ &= \ \mathbb{E}_W \left [\langle\psi_{\text{WP}}|(H_0 + H_W)^2|\psi_{\text{WP}}\rangle \ - \ (\langle\psi_{\text{WP}}|H_0 + H_W|\psi_{\text{WP}}\rangle)^2  \right ] \nonumber \\
&= \langle\psi_{\text{WP}}|H_0^2|\psi_{\text{WP}}\rangle - (\langle\psi_{\text{WP}}|H_0|\psi_{\text{WP}}\rangle)^2 +\mathbb{E}_W \left [\langle\psi_{\text{WP}}|(H_W)^2|\psi_{\text{WP}}\rangle \ - \ (\langle\psi_{\text{WP}}|H_W|\psi_{\text{WP}}\rangle)^2 \right ]  \ , 
\end{align}
In the second line, all terms linear in $H_W$ have been dropped as they average to zero.

First, consider the $H_0$ terms. 
Wavepackets are superpositions of momentum eigenstates that are also energy eigenstates with dispersion relations $H_0|\psi_k\rangle = E(k) |\psi_k\rangle$.
The energy without disorder is,
\begin{align}
\langle \psi_{\text{WP}}|H_0|\psi_{\text{WP}}\rangle \ &= \ {\cal N}^2\sum_k \sum_{k'} e^{i k' x_0}e^{-ikx_0}e^{-(k-k_0)^2/(4\sigma_p^2)}e^{-(k'-k_0)^2/(4\sigma_p^2)}E(k)\langle \psi_{k'} |\psi_k\rangle \nonumber \\
&= \ {\cal N}^2\sum_k e^{-(k-k_0)^2/(2\sigma_p^2)}E(k) \nonumber \\
&= \ {\cal N}^2\sum_{n=-N/2}^{N/2} e^{-4\pi^2(m-n)^2/(2\sigma_p^2 N^2)}E\left (\frac{2\pi n}{N} \right ) \nonumber \\
&= \ {\cal N}^2\sum_{l=-N/2-m}^{N/2-m} e^{-l^2/(2\tilde{\sigma}_p^2 )}E\left (\frac{2\pi (l+m)}{N} \right )  \ ,
\end{align}
where we have defined $k_0 = 2\pi m/N$ and $\tilde{\sigma}_p 2\pi/N= \sigma_p$. 
In the second line we used the orthonormality condition $\langle \psi_{k'}|\psi_k\rangle = \delta_{k,k'}$ and in the third line we assumed $N$ is odd and $N/2$ rounds down. 
We are interested in the large $N$ limit so this does not matter.
In the fourth line we defined $l =n-m$.

Next, assume that the wavepacket has its support away from the edges of the Brillouin zone, i.e. $k=\pi$,
\begin{align}
\frac{(\frac{N}{2}\pm m)^2}{2\tilde{\sigma}_p^2} \ = \ \frac{(\pi\pm k_0)^2}{2\sigma_p^2} \gg 1 \ .
\label{eq:assumption_1}
\end{align}
Additionally, let $\tilde{\sigma}_p = {\mathcal O}(N^0)$ which corresponds to a wavepacket that occupies a fixed fraction of the spatial volume.
With these assumptions, the range of the sum can be extended to $\pm\infty$ with exponentially small errors,
\begin{align}
\langle \psi_{\text{WP}}|H_0|\psi_{\text{WP}}\rangle \ &  \approx \ {\cal N}^2\sum_{l=-\infty}^{\infty}e^{-l^2/(2\tilde{\sigma}_p^2)}E\left (\frac{2\pi(l+m)}{N} \right ) \\
&= \ {\cal N}^2\sum_{l=-\infty}^{\infty}e^{-l^2/(2\tilde{\sigma}_p^2)}\left [E(k_0) \ + \ \frac{2\pi^2 l^2}{N^2} E''(k_0)  \ + \ {\mathcal O}(l^4/N^4)\right ]  \ . 
\end{align}
In the second line we have Taylor expanded the energy about $l/N=0$ and dropped the terms that are odd in $l$ as they are zero.
This Taylor expansion converges due to the Gaussian prefactor and $N\gg \tilde{\sigma}_p$ which is guaranteed by Eq.~\eqref{eq:assumption_1}.

It is convenient to define the moments of the Gaussian,
\begin{align}
M_n(\tilde{\sigma}_p) \ = \ \sum_{l=-\infty}^{\infty}l^n\,e^{-l^2/(2\tilde{\sigma}_p^2)} \ .
\end{align}
The normalization of $|\psi_{\text{WP}}\rangle$ fixes ${\cal N}^2 = 1/M_0$ following similar steps as above.
Putting everything together, the energy without disorder is
\begin{align}
\langle \psi_{\text{WP}}|H_0|\psi_{\text{WP}}\rangle \ = \ E(k_0) \ + \ \frac{2\pi^2}{N^2} \frac{M_2(\tilde{\sigma}_p)}{M_0(\tilde{\sigma}_p)}E''(k_0) + {\cal O}(1/N^4)
\end{align}

The expectation value of the square of the Hamiltonian without disorder follows similarly,
\begin{align}
\langle \psi_{\text{WP}} |H_0^2 |\psi_{\text{WP}}\rangle \ &\approx \ \ {\cal N}^2\sum_{l=-\infty}^{\infty}e^{-l^2/(2\tilde{\sigma}_p^2)}\left [E(k_0) \ + \frac{2 \pi l}{N} E'(k_0)\ +  \frac{2\pi^2 l^2}{N^2} E''(k_0)  \ + \ {\mathcal O}(l^3/N^3)\right ]^2 \nonumber \\
& = \ \ {\cal N}^2\sum_{l=-\infty}^{\infty}e^{-l^2/(2\tilde{\sigma}_p^2)}\left [E(k_0)^2 \ + \frac{4 \pi^2 l^2}{N^2} E'(k_0)^2\ +  \frac{4\pi^2 l^2}{N^2} E''(k_0)E(k_0)  \ + \ {\mathcal O}(l^4/N^4)\right ] \nonumber \\ 
& = \ E(k_0)^2 \ + \ \frac{4 \pi^2}{N^2}\frac{M_2(\tilde{\sigma}_p)}{M_0(\tilde{\sigma}_p)}\left [E'(k_0)^2 + E''(k_0)E(k_0) \right ] \ + \ {\mathcal O}(1/N^4)\ .
\end{align}

Now, the terms with disorder.
The square of the disorder Hamiltonian is,
\begin{align}
\mathbb{E}_W \left( \langle \psi_{\text{WP}}|H_W^2|\psi_{\text{WP}}\rangle \right ) \ &= \ \mathbb{E}_W \left( \langle \psi_{\text{WP}}|\sum_{i=0}^{N-1}W_in_i\sum_{j=0}^{N-1}W_j n_j|\psi_{\text{WP}}\rangle \right ) \nonumber \\
&= \ \frac{W^2}{3}\left( \langle \psi_{\text{WP}}|\sum_{i=0}^{N-1}n_i^2 |\psi_{\text{WP}}\rangle \right ) \nonumber \\
&= \ \frac{W^2}{3}N_e \ . 
\end{align}
In the second line we used $\mathbb{E}_W(\langle W_i W_j \rangle) = \delta_{i,j}W^2/3$ and in the third line we have used $n_i^2 = n_i$ and defined the total particle number as $N_e$.
The expectation of the disorder Hamiltonian squared is,
\begin{align}
\mathbb{E}_W \left [ \left( \langle \psi_{\text{WP}}|H_W|\psi_{\text{WP}}\rangle \right )^2 \right ] \ &= \ \mathbb{E}_W \left( \langle \psi_{\text{WP}}|\sum_{i=0}^{N-1}W_in_i|\psi_{\text{WP}}\rangle \langle \psi_{\text{WP}}|\sum_{j=0}^{N-1}W_j n_j|\psi_{\text{WP}}\rangle \right ) \nonumber \\
&= \ \frac{W^2}{3}\sum_{i=0}^{N-1} \langle \psi_{\text{WP}}|n_i |\psi_{\text{WP}}\rangle^2  \ . 
\end{align}

Generalizing to $D$ dimensions, it is found that the energy variance is,
\begin{align}
(\Delta E)^2 \ = \ \frac{4 \pi^2}{N^{2/D}}\frac{M_2(\tilde{\sigma}_p)}{M_0(\tilde{\sigma}_p)}\sum _{i=0}^{D-1}\left ( \partial_{k_{i}} E(\vec{k})\right )^2\bigg \rvert_{\vec{k} = \vec{k}_0} \ + \ \frac{W^2}{3}\left (N_e \ - \  \sum_{i=0}^{N-1} \langle \psi_{\text{WP}}|n_i |\psi_{\text{WP}}\rangle^2\right ) \ . 
\end{align}
The first term scales as ${\mathcal O}(N^{-2/D})$; i.e. as the inverse square of the side-length of the volume.
The scaling of the second term can be determined by noting that outside of the wavepacket the system is locally indistinguishable from the ground state which satisfies $\langle \psi_{\text{g.s.}}|n_i|\psi_{\text{g.s.}}\rangle=N_e/N$.
This implies that the second term scales as as ${\mathcal O}(W^2 N_e)$ provided the wavepacket is supported on at most a constant fraction of the volume.
For comparison, consider a random computational basis state with fixed $N_e$. The energy variance with respect to $H_0$ scales as ${\mathcal O}(N_e)$ while the variance with respect to $H_W$ is zero since it is an eigenstate.
As expected, wavepackets have the largest improvement in energy resolution over product states for weak disorder.

For a single particle $N_e=1$, the energy variance can be simplified to,
\begin{align}
(\Delta E_{N_e=1})^2 \ = \ \frac{16 \pi^2}{N^{2/D}}\frac{M_2(\tilde{\sigma}_p)}{M_0(\tilde{\sigma}_p)}\sum _{i=1}^{D}\sin^2{(k_{0,i})} \ + \ \frac{W^2}{3}\left ( 1 - \frac{\sum_{m=0}^{N-1}(c_m)^4}{\left (\sum_{m=0}^{N-1}(c_m)^2 \right )^2}\right ) \ .
\label{eq:DE_n1}
\end{align}
The $c_m$ are the coefficients of the wavepacket wavefunction in position space.
The ratio in the disorder term is ${\mathcal O}(1/N)$ as will be shown in the following.

For notational simplicity we again return to one dimension. 
The single-particle wavepacket in position space with $x_0=0$ is,
\begin{align}
|\psi_{\text{WP}}\rangle \ &= \ {\cal N}/\sqrt{N} \sum_{l=-N/2-m}^{N/2-m} e^{-l^2/(4\tilde{\sigma}_p^2)} \sum_{n=0}^{N-1} e^{\frac{2\pi}{N} i(l+m)n}|e_n\rangle \nonumber \\
& \approx \ {\cal N}/\sqrt{N} \sum_{n=0}^{N-1}e^{i k_0 n}\sum_{l=-\infty}^{\infty} e^{-l^2/(4\tilde{\sigma}_p^2)}  e^{\frac{2\pi}{N} iln}|e_n\rangle \nonumber \\
& = \ {\cal N}'\sum_{n=0}^{N-1}e^{i k_0 n}\sum_{l=-\infty}^{\infty} e^{-4 \pi^2\tilde{\sigma}_p^2(n/N + l)^2}  |e_n\rangle \nonumber \\
& \equiv \ {\cal N}'\sum_{n=0}^{N-1}e^{i k_0 n}c_n  |e_n\rangle \ ,
\end{align}
where ${\cal N}'$ is a new normalization constant.
In line one we used Eq.~\eqref{eq:freePlaneWave} and in line two the same steps as above have been used.
The third line applies the Poisson resummation formula which improves the convergence of the sum.
The normalization condition is,
\begin{align}
({\cal N}')^2 = \left ( \sum_{n=0}^{N-1}(c_n)^2\right )^{-1} \quad , \quad c_n \equiv \sum_{l=-\infty}^{\infty} e^{-4 \pi^2\tilde{\sigma}_p^2(n/N + l)^2} \approx  e^{-4 \pi^2\tilde{\sigma}_p^2(n/N)^2}
\left [1+e^{-4\pi^2 \tilde{\sigma}_p^2(1-2n/N)} \right ]
\end{align}
where the approximation just keeps the $l=0,-1$ terms in the sum. This is an exponentially good approximation provided that the spatial width of the Gaussian wavepacket is less than the system size, roughly corresponding to $\tilde{\sigma}_p >1/2$.
Using this approximation, the sums in Eq.~\eqref{eq:DE_n1} can be evaluated as,
\begin{align}
    \sum_{n=0}^{N-1}(c_n)^d  \ &= \ \sum_{k=0}^{d}\binom{d}{k}e^{-4\pi^2 \tilde{\sigma}_p^2 k(1-k/d)}\sum_{n=0}^{N-1}e^{-4\pi^2 \tilde{\sigma}_p^2d(n/N-k/d)^2} \nonumber \\
    & \approx \ N\sum_{k=0}^{d}\binom{d}{k}e^{-4\pi^2 \tilde{\sigma}_p^2 k(1-k/d)}\int_0^1 dx\,  e^{-4\pi^2 \tilde{\sigma}_p^2d(x-k/d)^2} \ .
\end{align}
The integral in the second line is a good approximation for large $N$ since the sum varies smoothly due to $4\pi^2 \tilde{\sigma}_p d/N^2\ll 1$.
This integral can be explicitly evaluated in terms of error functions, but the important observation is that $\sum_{n=0}^{N-1}(c_n)^d$ is ${\mathcal O}(N)$ for the $d=2,4$ that are relevant.
The extension to arbitrary dimensions is straightforward and does not change the scaling with $N$.
As a results, the ratio in the disorder term in Eq.~\eqref{eq:DE_n1} scales as $\mathcal{O}(1/N)$.

\section{Quantum Circuits}
\label{a:qcircs}
\noindent
We now discuss in more detail the circuits used to prepare the single-particle wavepackets in Section~\ref{s:simulations} and $|W(k_0)\rangle$ in Section~\ref{s:halffilling}.
Effective, high-fidelity preparation of the wavepacket defined in Eq.~\eqref{eq:2dWP} on the lattice
\begin{equation}
    |\psi_{\text{WP}}\rangle = \mathcal{N} \sum_{x}\sum_{y} c_{x, y} |e_{x, y}\rangle
    \label{eq:2DWP2}
\end{equation}
is important because it initializes the simulation.
Deviations from the expected wavepacket introduce uncertainty in measurements of the IPR and impede identification of the finite-size mobility edge.
For $c_{x,y} = 1/\sqrt{N}$ the single-particle wavepacket is $|W_N\rangle$, the $N$-qubit W state~\cite{Dur:2000zz}.
The preparation of $|W_N\rangle$ has been extensively explored due to its usage in quantum sensing and communication~\cite{Joo:2003scl,Agrawal:2006cbk,LIU20113160}.
Many of these protocols can be adapted to the more general case of wavepackets. 
To explore their noise robustness, we will focus on the preparation of $|W_N\rangle$.

Prior works have presented W state preparation circuits tailored to a target qubit connectivity.
For linear connectivity the depth scales as $\mathcal{O}(N)$~\cite{Cruz_2019}, for 2D grid connectivity the depth improves to $\mathcal{O}(\sqrt{N})$~\cite{Yuan:2025dqo,Bartschi:2022dicke} and for all-to-all connectivity the depth scales as $\mathcal{O}(\log{N})$~\cite{Cruz_2019} or $\mathcal{O}(1)$ with ancillas~\cite{Vasconcelos:2026sqs}.
Additionally, recent methods~\cite{Buhrman:2023rft,Piroli:2024mcm,Yu:2024szp,Farrell:2025nkx} have begun leveraging mid-circuit measurement and feedforward (MCM-FF) to reduce the circuit depth at the cost of an increased shot overhead and/or use of ancillas.
Notably, Ref.~\cite{Piroli:2024mcm} developed a protocol for preparing W states in constant depth using ancillas that were later removed in Ref.~\cite{Farrell:2025nkx}.
Due to their constant depth, these approaches are expected to be optimal for asymptotically large systems.

\begin{figure}
    \centering
    \includegraphics[width=0.8\linewidth]{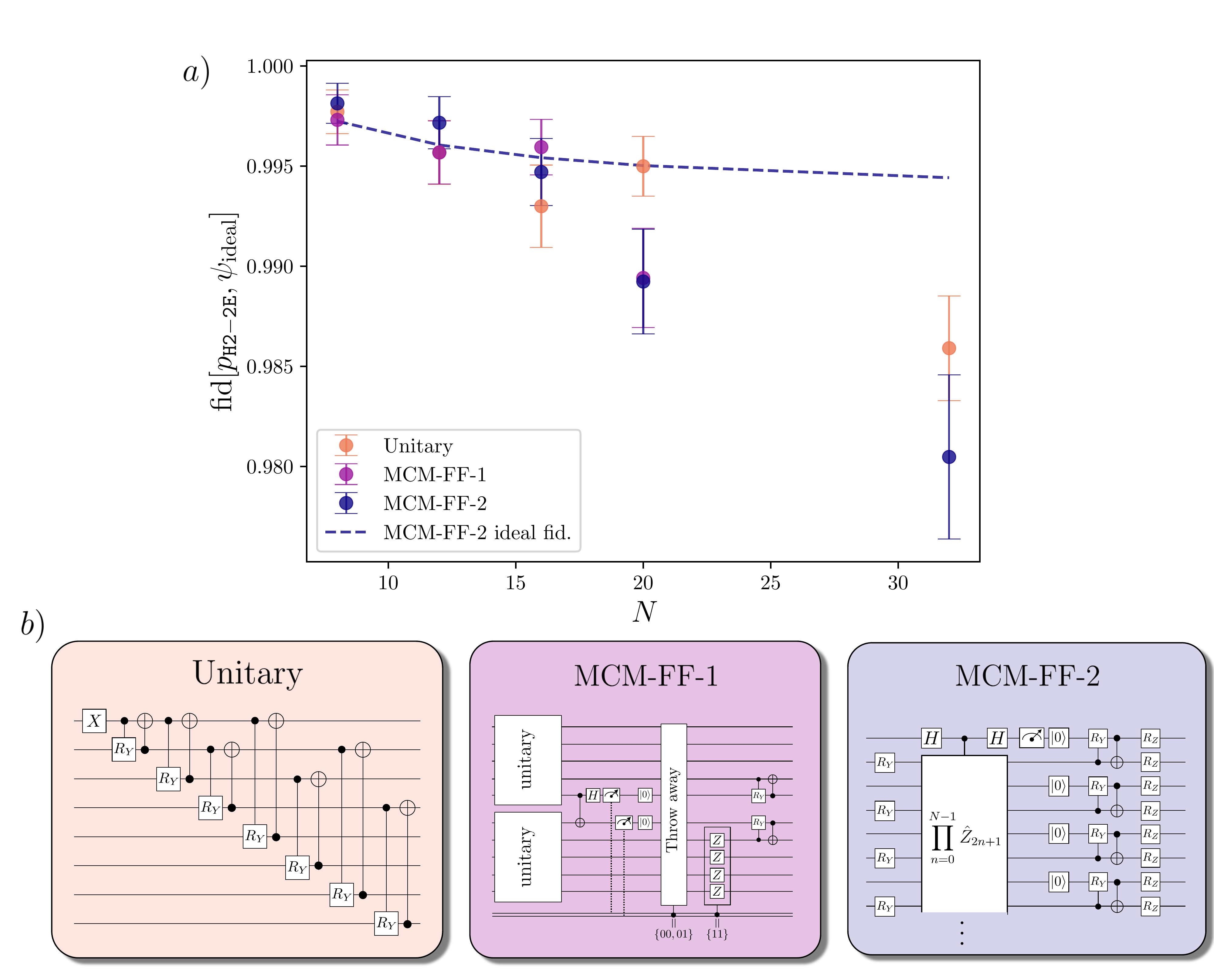}
    \caption{\textit{Wavepacket preparation circuits and noisy W state preparation results.} a) The classical fidelity defined in Eq.~\eqref{eq:classical_fidelity} of W states prepared on  Quantinuum's noisy {\tt H2-Emulator} and {\tt H2-1E} classical emulators.
    Three W state preparation circuits were tested and are shown in the bottom panels.
    The total number of shots was adjusted so that each method retained 1,350 shots after post-selection.
    The MCM-FF-2 method prepares an approximate W state with an ideal fidelity that is shown by the dashed blue line and tabulated in Table~\ref{tab:mcmff2_ideal_metrics}.
    b) The circuit structure for each state preparation method.
    }
    \label{fig:WP_prep_comparisons}
\end{figure}

In this appendix we compare three different W state preparation protocols using the Quantinuum {\tt H2} noisy emulators.\footnote{For $N\leq 20$ we used the {\tt H2-Emulator} that can be run locally and for larger $N$ we used ${\tt H2-2E}$ accessed through Quantinuum Nexus~\cite{quantinuum_nexus}. 
}
The corresponding quantum circuits are shown in Fig.~\ref{fig:WP_prep_comparisons}b) and more detail can be found in Ref.~\cite{Farrell:2025nkx}.
The $\log N$ depth unitary method that uses the all-to-all connectivity available on Quantinuum's devices is found to perform the best and was used in Section~\ref{s:simulations} to initialize single-particle wavepackets on {\tt H2-2}.
\begin{enumerate}
    \item \textbf{Unitary circuit}: The $\log{N}$ depth W state preparation circuit presented in Ref.~\cite{Cruz_2019}.
    This method has no mid-circuit measurements and therefore has no shot overhead. It also requires the fewest number of two-qubit gates of the methods tested.
    
    \item \textbf{MCM-FF-1}: This method prepares smaller W states on each half of the qubit register using the unitary circuit, and then fuses them into a large W state with a Bell measurement~\cite{Farrell:2025nkx}.
    The measurement is followed by a feedforward operation that prepares the target W state successfully with probability $p_{\text{success}}=1/2$.
    The circuit depth and gate counts are similar to the unitary circuit, but it incurs a $2\times$ shot overhead due to post-selection.

    \item \textbf{MCM-FF-2 circuit}: This is the constant depth ancilla-free method presented in Ref.~\cite{Farrell:2025nkx} that extends Ref.~\cite{Piroli:2024mcm}.
    This strategy takes as input a parameter $\delta$ which sets the fidelity of the prepared state as well as the shot overhead.
    The inverse of the success probability gives the shot overhead and is,
    \begin{align}
        p_{\text{success}} \ 
        = \  \frac{1}{2} - \frac{1}{2} \left( 1 - \frac{4\delta}{N} \right)^{N / 2} \ .
        \label{eq:mcmff2_psuccess}
    \end{align}
    Additionally, $\delta$ sets the fidelity of the prepared state,
    \begin{align}
        |\langle W_N|\psi_{\text{MCM-FF-2}}\rangle|^2\ = \ \frac{1}{p_{\text{success}}} \delta \left(1 - \frac{2\delta}{N} \right)^{N / 2 - 1} \ ,
        \label{eq:mcmff2_ideal_fid}
    \end{align}
    where $|\psi_{\text{MCM-FF-2}}\rangle$ is the state prepared using MCM-FF-2 on an ideal quantum device.
    Of the three methods, the MCM-FF-2 circuit requires the largest number of gates and (parallelizable) measurement operations, but has the best asymptotic two-qubit gate depth scaling.
\end{enumerate}

\begin{table}[t]
\centering
\renewcommand{\arraystretch}{1.4}
\begin{tabularx}{0.6\linewidth}{|c||Y|Y|Y|Y|Y|Y|} 
\hline
\makecell{$N$}  & \makecell{8} & \makecell{12} & \makecell{16}  & \makecell{20} & \makecell{32} \\
\hline\hline
$|\langle W|\psi_{\text{MCM-FF-2}}\rangle|^2$ & 0.9972 & 0.9961 & 0.9954 & 0.9950 & 0.9944 \\ \hline
$p_{\text{success}}$ & 0.1720 & 0.1695 & 0.1683 & 0.1676 & 0.1665 \\ \hline
\end{tabularx}
\renewcommand{\arraystretch}{1}
\caption{The ideal fidelity and $p_{\text{success}}$ of the W state prepared using the MCM-FF-2 method for the system sizes run on the {\tt H2} noisy emulators. These quantities are computed from Eq.~\eqref{eq:mcmff2_ideal_fid} and Eq.~\eqref{eq:mcmff2_psuccess} using $\delta=0.2$. 
}
\label{tab:mcmff2_ideal_metrics}
\end{table}

To gauge the success of each method, we calculate the classical fidelity,
\begin{align}
\text{Fid}[p_{ {\tt Emulator}},\psi_{\text{ideal}}] \ = \ \sum_{n=0}^{N-1} \sqrt{p_{ {\tt Emulator}}(n) |\langle e_{n}|\psi_{\text{ideal}}\rangle |^2} \ ,
\label{eq:classical_fidelity}
\end{align}
where $|\psi_{\text{ideal}}\rangle = |W_N\rangle$ and $p_{ {\tt Emulator}}(n)$ is the probability of measuring $|e_{n}\rangle$ determined from the noisy emulator without any error mitigation.
For simplicity, we have flattened the 2D index $c_{x,y}$ with $c_n = c_{L_x y +x}$.
This metric is computing the fidelity between the probability distributions in the computational basis.
As such it is insensitive to relative phases in the wavefunctions that are important for determining, e.g., the energy density.
The fidelity obtained from the noisy emulator for the three methods for various system sizes are shown in Fig.~\ref{fig:WP_prep_comparisons}a).
The number of shots has been adjusted so that all three methods have the same number of successful shots, therefore the uncertainty coming from shot noise is the same. 
Issues with Quantinuum Nexus prevented MCM-FF-1 from being successfully run for $N=32$.
Surprisingly, all the methods are essentially consistent with each other within error bars. 
Thus, due to the shot overhead incurred by MCM-FF-1 and MCM-FF-2, we conclude that the unitary circuit currently performs the best.
For asymptotically large system sizes, we expect that the circuit depth will become a stronger determining factor in deciding the best state preparation method to use, giving MCM-FF-2 an advantage.

Based on the above considerations, we chose the unitary circuit in Fig.~\ref{fig:WP_prep_comparisons}b) to prepare wavepackets on Quantinuum's {\tt H2-2} quantum computer.
The wavepackets are specified by the $c_n = c_{L_x y+x}$ in Eq.~\eqref{eq:2dWP}.
The $R_Y$ rotation angles in the unitary circuit depend on the $c_n$ through a system of linear equations~\cite{Farrell:2025nkx}.
For complex $c_n$ there is an additional step needed to prepare the target wavepacket.
A final layer of single qubit $R_Z$ rotations with rotation angles $\prod_n e^{-i \arctan{\left (\frac{\text{Im}[c_n]}{\text{Re}[c_n]} \right)} \hat{Z}_n /2}$ is applied to put the proper phases into the wavepacket.
Sample code for computing the coefficients $c_n$ and generating the quantum circuits that prepare wavepackets and W states using the unitary, MCM-FF-1 and MCM-FF-2 methods are available at the Github repository \href{https://github.com/malee4/gaussian-WP-prep}{https://github.com/malee4/gaussian-WP-prep}.

\section{Additional details on quantum simulations}
\label{a:qsim_params}
\begin{figure}
    \centering
    \includegraphics[width=0.75\linewidth]{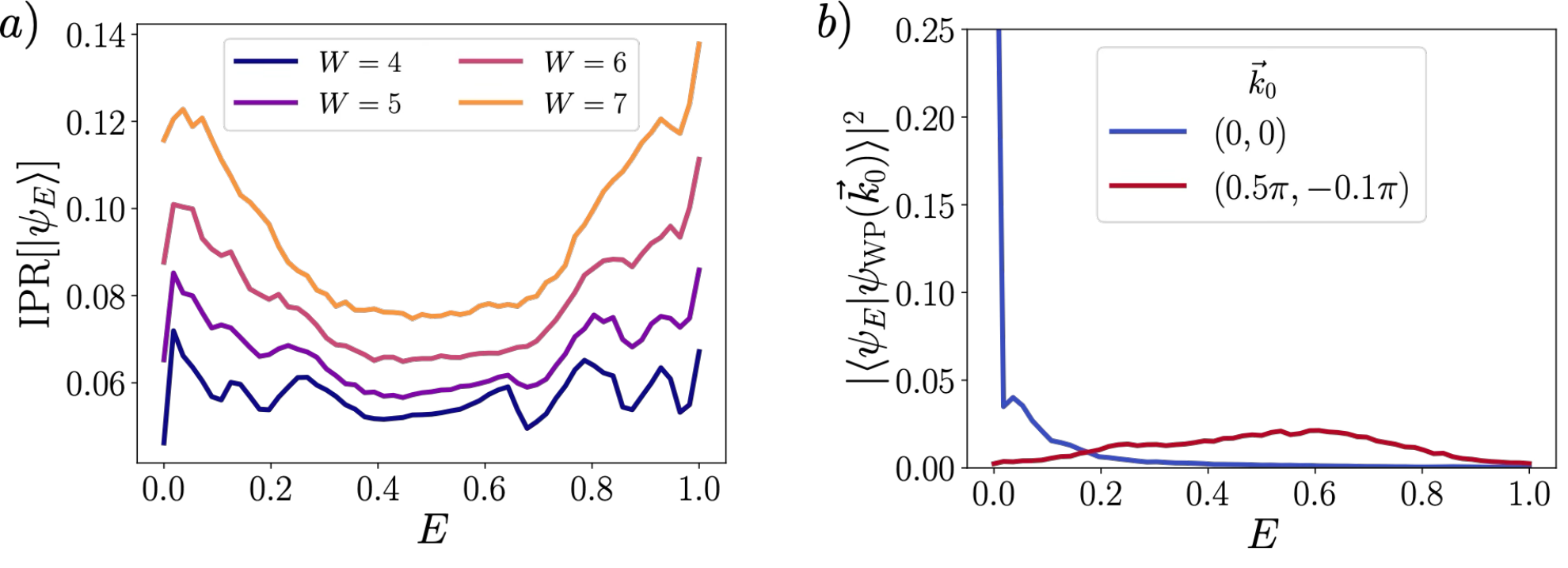}
    \caption{a) The IPR of the energy eigenstates for a selection of disorder strengths $W$.
    b) The overlap of a low-energy wavepacket with $\vec{k}_0=(0,0)$ and a high-energy wavepacket with $\vec{k}_0=(0.5\pi,-0.1\pi)$ onto energy eigenstates with $W=6$. The wavepackets have $\vec{\sigma}_p = (0.3,0.35)$.
    Both plots are for a $8\times7$ lattice and have been averaged over 2000 disorder realizations.
    Energies have been rescaled to lie in the interval $E\in[0,1]$.
    }
    \label{fig:Energy_dependent_WP_8x7}
\end{figure}

\begin{figure}
    \centering
    \includegraphics[width=\linewidth]{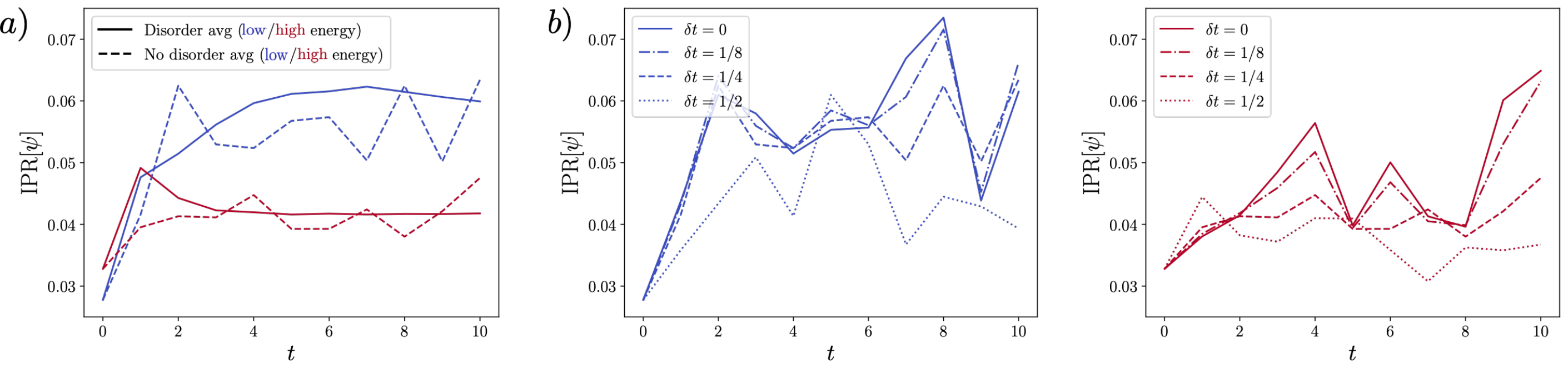}
    \caption{a) The IPR using the disorder instance $W_i$ employed in the quantum simulations (dashed lines) compared to the disorder averaged over 2000 instances (solid lines).
    b) The IPR for a variety of Trotter step sizes $\delta t$, shown both for the low-energy wavepacket (blue) and high-energy wavepacket (red). 
    No truncation of the initial wavepacket was used in either figure.}
    \label{fig:TrotterError_8x7}
\end{figure}

\noindent
This appendix outlines the motivation behind some of the parameters chosen for the quantum simulations performed on {\tt {\tt H2-2}} in Section~\ref{s:simulations}.
\begin{itemize}
    \item Disorder strength $W$: A larger $W$ leads to increased contrast between the localization of low- and high-energy eigenstates. 
    This is illustrated in Fig.~\ref{fig:Energy_dependent_WP_8x7}a) which shows the IPR as a function of energy for a range of $W$.
    A larger $W$ also increases the energy variance of the initial wavepacket state, limiting the ability to target energy-dependent localization. 
    $W=6$ was chosen to balance these effects. 
    \item Wavepacket momentum $\vec{k}_0$: The low energy wavepacket momentum was chosen to be $\vec{k}_0=(0,0)$ as this always has the lowest energy. 
    The momentum of the high-energy wavepacket was chosen to maximize the overlap onto eigenstates that are the most delocalized.
    From Fig.~\ref{fig:Energy_dependent_WP_8x7}a) it is seen that this approximately corresponds to states with $E\in[0.25,0.75]$. The critical energies of the finite-size mobility edge in this system are therefore approximately $E_c=0.25$ and $E_c=0.75$.  
    The momentum $\vec{k}_0=(0.5\pi,-0.1\pi)$ was found to maximize the overlap onto states in this energy interval.
    The overlaps of the low and high-energy wavepackets onto the energy eigenstates are shown in Fig.~\ref{fig:Energy_dependent_WP_8x7}b).
    \item Trotter step size $\delta t$: The Trotter step size should balance Trotter errors against the number of two-qubit gates required to reach a target simulation time.
    $\delta t=1/4$ was chosen from examining the convergence with $\delta t$ of the IPR for times $t\in[0,1,2]$, shown in Fig.~\ref{fig:TrotterError_8x7}b). 
    Over this time interval the Trotter errors are negligible.
    \item  The site-dependent disorder instance $W_i$: Statements about energy-dependent localization are only precise when averaged over many disorder realizations.
    However, due to limited quantum resources, our quantum simulations only used one disorder instance.
    This disorder instance was chosen to have a time-dependent IPR that is close to the average as shown in Fig.~\ref{fig:TrotterError_8x7}a).
\end{itemize}
%

\section{Additional quantum results}
\label{a:qsim_details}
\begin{figure}
    \centering
    \includegraphics[width=\linewidth]{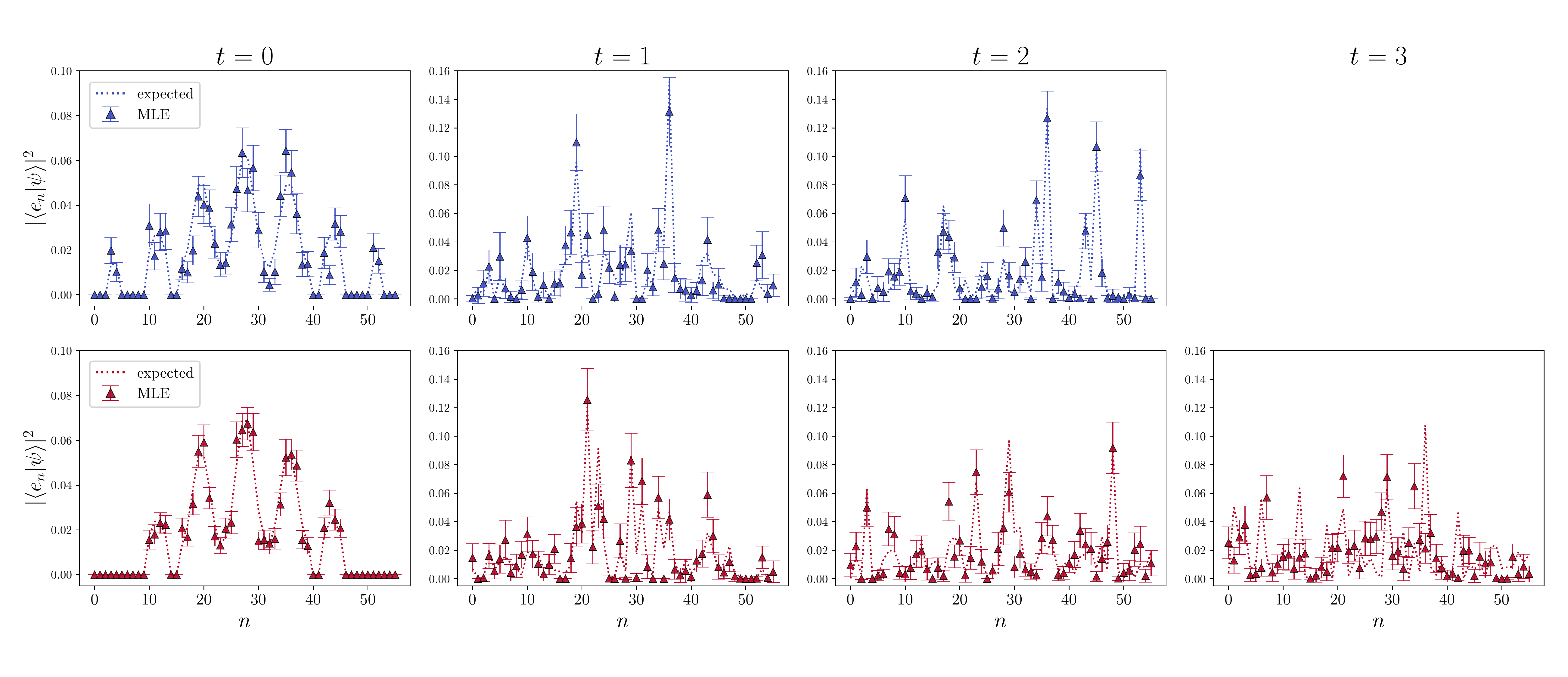}
    \caption{The probability of measuring a particle to be at site $n=x+L_x y$ obtained from ideal classical simulations (dotted lines) and {\tt H2-2} (data) post-processed with MLE error mitigation.
    The upper (lower) row of plots corresponds to the dynamics of a low (high) energy wavepacket.}
    \label{fig:all_results}
\end{figure}

\noindent
The probability distributions obtained from {\tt H2-2} using max-likelihood estimation (MLE) and post-selection (PS) error mitigation can be compared quantitatively with the classical fidelity $\text{Fid}[p_{ {\tt H2-2}},\psi_{\text{ideal}}]$ defined in Eq.~\eqref{eq:classical_fidelity}.
This compares the classically computed noiseless wavefunction $|\psi_{\text{ideal}}\rangle$ to $p_{{\tt {\tt H2-2}}}(x,y)$, the probability of measuring $|e_{x,y}\rangle$ determined from {\tt H2-2} after error mitigation.
The classical infidelities of the low- and high-energy wavepackets are given in Table~\ref{tab:quantum_result_fidelities}.
Due to shot noise, the ideal, i.e. noiseless, fidelity is not equal to 1, and has been included as a reference.
The fidelities obtained with MLE are consistently higher than with PS, with the improvement becoming more pronounced at later times due to the lower post-selection rates.
\begin{table}[t]
\centering
\renewcommand{\arraystretch}{1.4}
\begin{tabularx}{0.6\linewidth}{|c||Y|Y|Y|Y|Y|Y|} 
\hline
& \multicolumn{3}{c|}{Low-energy wavepacket fidelity} &\multicolumn{3}{c|}{High-energy wavepacket fidelity} \\
\hline
\makecell{$t$}  & \makecell{PS} & \makecell{MLE} & \makecell{Ideal}  & \makecell{PS} & \makecell{MLE} & \makecell{Ideal}  \\
\hline\hline
0 & 0.950(11) & 0.955(10) & 0.981(5) & 0.982(4) & 0.982(4) & 0.992(2)\\ \hline
1 & 0.691(47) & 0.815(28) & 0.935(13) & 0.740(29) & 0.810(23) & 0.928(13)\\ \hline
2 & 0.693(32) & 0.800(22) & 0.954(8) & 0.636(46) & 0.780(30) & 0.958(9)\\ \hline
3 & -- & -- & -- & 0.302(44) & 0.535(37) & 0.963(7) \\ \hline
\end{tabularx}
\renewcommand{\arraystretch}{1}
\caption{The classical fidelity, defined in Eq.~\eqref{eq:classical_fidelity} of the exact state with the measured state on {\tt H2-2} after error mitigation.
Results are shown for the state obtained after post-selection (PS) on $N_e=1$, error mitigation with max-likelihood estimation (MLE) and the ideal results only assuming shot noise. 
The total number of shots for each simulation time are given in Table~\ref{tab:PS_rates}.}
\label{tab:quantum_result_fidelities}
\end{table}
\begin{table}[t]
\centering
\renewcommand{\arraystretch}{1.4}
\begin{tabularx}{0.6\linewidth}{|c||c|Y|Y|c|Y|Y|} 
\hline
& \multicolumn{3}{c|}{Low-energy IPR} &\multicolumn{3}{c|}{High-energy IPR} \\
\hline
$t$  & PS & MLE & Ideal  & PS & MLE & Ideal \\
\hline\hline
0 & $0.040(2)$ & $0.039(2)$ & 0.035  & $0.042(1)$ & $0.042(1)$ & 0.040 \\ \hline
1 & $0.069 (12)$ & $0.059(6)$ & 0.054  & $0.057(6)$ & $0.055(6)$ & 0.049  \\ \hline
2 & $0.091(9)$ & $0.064(4)$ &  0.062 & $0.052(7)$ & $0.043(3)$ & 0.041 \\ \hline
3 & -- & -- & 0.053 & $0.061 (10)$ & $0.039(2)$ & 0.041 \\
\hline
\end{tabularx}
\renewcommand{\arraystretch}{1}
\caption{The IPR, defined in Eq.~\eqref{eq:IPR}, of the state obtained on {\tt {\tt H2-2}} for  a selection of simulation times $t$.
Results are shown for the state obtained after post-selection (PS) on $N_e=1$, error mitigation with max-likelihood estimation (MLE) and the ideal result from classical simulation.}
\label{tab:quantum_result_IPR}
\end{table}
A more fine-grained analysis of the MLE mitigated probability distributions is shown in 
Fig.~\ref{fig:all_results} which gives the probability density $p_{x,y}$ for all simulation times.
The probability distributions have been flattened to one dimension with the index $n =  L_xy+x$. 
With MLE error mitigation, even small features of the ideal probability distribution are reproduced.
Statistically significant deviations away from the ideal probability density become more frequent at later times due to the increased two-qubit gate counts.
Interestingly, these systematic error cancel out in the computation of the IPR, which remains consistent with the ideal IPR even at $t=3$ as shown in Table~\ref{tab:quantum_result_IPR}.

\clearpage

\bibliography{bibi}

@article{anderson1958absence,
  title={Absence of diffusion in certain random lattices},
  author={Anderson, Philip W},
  journal={Physical review},
  volume={109},
  number={5},
  pages={1492},
  year={1958},
  publisher={APS}
}

@article{li2025many,
  title={Many-body delocalization with a two-dimensional 70-qubit superconducting quantum simulator},
  author={Li, Tian-Ming and Sun, Zheng-Hang and Shi, Yun-Hao and Bao, Zhen-Ting and Wang, Yong-Yi and Zhang, Jia-Chi and Liu, Yu and Deng, Cheng-Lin and Yu, Yi-Han and Liu, Zheng-He and others},
  journal={arXiv preprint arXiv:2507.16882},
  year={2025}
}

@article{Billy_2008,
   title={Direct observation of Anderson localization of matter waves in a controlled disorder},
   volume={453},
   ISSN={1476-4687},
   url={http://dx.doi.org/10.1038/nature07000},
   DOI={10.1038/nature07000},
   number={7197},
   journal={Nature},
   publisher={Springer Science and Business Media LLC},
   author={Billy, Juliette and Josse, Vincent and Zuo, Zhanchun and Bernard, Alain and Hambrecht, Ben and Lugan, Pierre and Clément, David and Sanchez-Palencia, Laurent and Bouyer, Philippe and Aspect, Alain},
   year={2008},
   month=jun, pages={891–894} }

@article{Luschen:2017cir,
    author = {L{\"u}schen, Henrik P. and Scherg, Sebastian and Kohlert, Thomas and Schreiber, Michael and Bordia, Pranjal and Li, Xiao and Sarma, S. Das and Bloch, Immanuel},
    title = "{Exploring the Single-Particle Mobility Edge in a One-Dimensional Quasiperiodic Optical Lattice}",
    eprint = "1709.03478",
    archivePrefix = "arXiv",
    primaryClass = "cond-mat.quant-gas",
    doi = "10.1103/PhysRevLett.120.160404",
    journal = "Phys. Rev. Lett.",
    volume = "120",
    pages = "160404",
    year = "2018"
}

@article{PhysRevLett.126.040603,
  title = {Interactions and Mobility Edges: Observing the Generalized Aubry-Andr\'e Model},
  author = {An, Fangzhao Alex and Padavi\ifmmode \acute{c}\else \'{c}\fi{}, Karmela and Meier, Eric J. and Hegde, Suraj and Ganeshan, Sriram and Pixley, J. H. and Vishveshwara, Smitha and Gadway, Bryce},
  journal = {Phys. Rev. Lett.},
  volume = {126},
  issue = {4},
  pages = {040603},
  numpages = {6},
  year = {2021},
  month = {Jan},
  publisher = {American Physical Society},
  doi = {10.1103/PhysRevLett.126.040603},
  url = {https://link.aps.org/doi/10.1103/PhysRevLett.126.040603}
}

@article{Guo:2019eke,
    author = "Guo, Qiujiang and others",
    title = "{Observation of energy-resolved many-body localization}",
    eprint = "1912.02818",
    archivePrefix = "arXiv",
    primaryClass = "quant-ph",
    doi = "10.1038/s41567-020-1035-1",
    journal = "Nature Phys.",
    volume = "17",
    number = "2",
    pages = "234--239",
    year = "2020"
}

@article{An:2018qfi,
    author = "An, Fangzhao Alex and Meier, Eric J. and Gadway, Bryce",
    title = "{Engineering a flux-dependent mobility edge in disordered zigzag chains}",
    eprint = "1705.09268",
    archivePrefix = "arXiv",
    primaryClass = "cond-mat.quant-gas",
    doi = "10.1103/PhysRevX.8.031045",
    journal = "Phys. Rev. X",
    volume = "8",
    pages = "031045",
    year = "2018"
}

@article{yao2023observation,
  title={Observation of many-body {F}ock space dynamics in two dimensions},
  author={Yao, Yunyan and Xiang, Liang and Guo, Zexian and Bao, Zehang and Yang, Yong-Feng and Song, Zixuan and Shi, Haohai and Zhu, Xuhao and Jin, Feitong and Chen, Jiachen and others},
  journal={Nature Physics},
  volume={19},
  number={10},
  pages={1459--1465},
  year={2023},
  publisher={Nature Publishing Group UK London}
}

@article{RevModPhys.57.287,
  title = {Disordered electronic systems},
  author = {Lee, Patrick A. and Ramakrishnan, T. V.},
  journal = {Rev. Mod. Phys.},
  volume = {57},
  issue = {2},
  pages = {287--337},
  numpages = {0},
  year = {1985},
  month = {Apr},
  publisher = {American Physical Society},
  doi = {10.1103/RevModPhys.57.287},
  url = {https://link.aps.org/doi/10.1103/RevModPhys.57.287}
}

@misc{Broers:2024eoa,
    author = "Broers, Lukas and Mathey, Ludwig",
    title = "{Exclusive-or encoded algebraic structure for efficient quantum dynamics}",
    eprint = "2404.09312",
    archivePrefix = "arXiv",
    primaryClass = "cond-mat.other",
    month = "4",
    year = "2024"
}

@misc{Begusic:2023jwi,
    author = "Begu{\v{s}}i{\'c}, Tomislav and Chan, Garnet Kin-Lic",
    title = "{Fast classical simulation of evidence for the utility of quantum computing before fault tolerance}",
    eprint = "2306.16372",
    archivePrefix = "arXiv",
    primaryClass = "quant-ph",
    month = "6",
    year = "2023"
}

@article{Deng:2017pen,
    author = "Deng, Dong-Ling and Li, Xiaopeng and Pixley, J. H. and Wu, Yang-Le and Sarma, S. Das",
    title = "{Logarithmic Entanglement Lightcone in Many-Body Localized Systems}",
    eprint = "1607.08611",
    archivePrefix = "arXiv",
    primaryClass = "cond-mat.dis-nn",
    doi = "10.1103/PhysRevB.95.024202",
    journal = "Phys. Rev. B",
    volume = "95",
    pages = "024202",
    year = "2017"
}

@article{Gong:2020dav,
    author = "Gong, Ming and others",
    title = "{Experimental characterization of the quantum many-body localization transition}",
    eprint = "2012.11521",
    archivePrefix = "arXiv",
    primaryClass = "quant-ph",
    doi = "10.1103/PhysRevResearch.3.033043",
    journal = "Phys. Rev. Res.",
    volume = "3",
    number = "3",
    pages = "033043",
    year = "2021"
}

@misc{chen2016universallogarithmicscramblingbody,
      title={Universal Logarithmic Scrambling in Many Body Localization}, 
      author={Yu Chen},
      year={2016},
      eprint={1608.02765},
      archivePrefix={arXiv},
      primaryClass={cond-mat.dis-nn},
      url={https://arxiv.org/abs/1608.02765}, 
}

@article{Huang_2016,
   title={Out‐of‐time‐ordered correlators in many‐body localized systems},
   volume={529},
   ISSN={1521-3889},
   url={http://dx.doi.org/10.1002/andp.201600318},
   DOI={10.1002/andp.201600318},
   number={7},
   journal={Annalen der Physik},
   publisher={Wiley},
   author={Huang, Yichen and Zhang, Yong‐Liang and Chen, Xie},
   year={2016},
   month=dec }

@misc{Rudolph:2025gyq,
    author = {Rudolph, Manuel S. and Jones, Tyson and Teng, Yanting and Angrisani, Armando and Holmes, Zo{\"e}},
    title = "{Pauli Propagation: A Computational Framework for Simulating Quantum Systems}",
    eprint = "2505.21606",
    archivePrefix = "arXiv",
    primaryClass = "quant-ph",
    month = "5",
    year = "2025"
}

@article{PhysRev.94.262,
  title = {Atomic Theory of the Two-Fluid Model of Liquid Helium},
  author = {Feynman, R. P.},
  journal = {Phys. Rev.},
  volume = {94},
  issue = {2},
  pages = {262--277},
  numpages = {0},
  year = {1954},
  month = {Apr},
  publisher = {American Physical Society},
  doi = {10.1103/PhysRev.94.262},
  url = {https://link.aps.org/doi/10.1103/PhysRev.94.262}
}

@article{Joo:2003scl,
    author = "Joo, Jaewoo and Park, Young-Jai and Oh, Sangchul and Kim, Jaewan",
    title = "{Quantum teleportation via a W state}",
    eprint = "quant-ph/0306175",
    archivePrefix = "arXiv",
    doi = "10.1088/1367-2630/5/1/136",
    journal = "New J. Phys.",
    volume = "5",
    number = "1",
    pages = "136",
    year = "2003"
}

@article{LIU20113160,
title = {An efficient protocol for the quantum private comparison of equality with W state},
journal = {Optics Communications},
volume = {284},
number = {12},
pages = {3160-3163},
year = {2011},
issn = {0030-4018},
doi = {https://doi.org/10.1016/j.optcom.2011.02.017},
url = {https://www.sciencedirect.com/science/article/pii/S0030401811001842},
author = {Wen Liu and Yong-Bin Wang and Zheng-Tao Jiang}
}

@misc{Li:2025sgo,
    author = "Li, Zhiyao and Illa, Marc and Savage, Martin J.",
    title = "{A Framework for Quantum Simulations of Energy-Loss and Hadronization in Non-Abelian Gauge Theories: SU(2) Lattice Gauge Theory in 1+1D}",
    eprint = "2512.05210",
    archivePrefix = "arXiv",
    primaryClass = "quant-ph",
    reportNumber = "IQuS@UW-21-115",
    month = "12",
    year = "2025"
}

@article{Alet_2018,
   title={Many-body localization: An introduction and selected topics},
   volume={19},
   ISSN={1878-1535},
   url={http://dx.doi.org/10.1016/j.crhy.2018.03.003},
   DOI={10.1016/j.crhy.2018.03.003},
   number={6},
   journal={Comptes Rendus. Physique},
   publisher={MathDoc/Centre Mersenne},
   author={Alet, Fabien and Laflorencie, Nicolas},
   year={2018},
   month=apr, pages={498–525} }

@misc{quantinuum_nexus,
  title={Quantinuum Nexus},
  url={https://nexus.quantinuum.com/},
  year={2024},
}

@misc{Vasconcelos:2026sqs,
    author = "Vasconcelos, Francisca and Joshi, Malvika Raj",
    title = "{Constant-Depth Unitary Preparation of Dicke States}",
    eprint = "2601.10693",
    archivePrefix = "arXiv",
    primaryClass = "quant-ph",
    month = "1",
    year = "2026"
}

@article{Puente:2024bxt,
    author = "Puente, Daniel Alcalde and Rizzi, Matteo",
    title = "{Learning Feedback Mechanisms for Measurement-Based Variational Quantum State Preparation}",
    eprint = "2411.19914",
    archivePrefix = "arXiv",
    primaryClass = "quant-ph",
    doi = "10.22331/q-2025-07-11-1792",
    journal = "Quantum",
    volume = "9",
    pages = "1792",
    year = "2025"
}

@misc{Niu:2024oxx,
    author = "Niu, Siyuan and Kokcu, Efekan and Mitra, Anupam and Szasz, Aaron and Hashim, Akel and Kalloor, Justin and de Jong, Wibe Albert and Iancu, Costin and Younis",
    title = "{AC/DC: Automated Compilation for Dynamic Circuits}",
    eprint = "2412.07969",
    archivePrefix = "arXiv",
    primaryClass = "quant-ph",
    month = "12",
    year = "2024"
}

@misc{Alam:2024mit,
    author = "Alam, Faisal and Clark, Bryan K.",
    title = "{Learning dynamic quantum circuits for efficient state preparation}",
    eprint = "2410.09030",
    archivePrefix = "arXiv",
    primaryClass = "quant-ph",
    reportNumber = "LA-UR-24-30605",
    month = "10",
    year = "2024"
}

@article{Yan:2024xev,
    author = "Yan, Yuxuan and Ma, Muzhou and Zhou, You and Ma, Xiongfeng",
    title = "{Variational LOCC-Assisted Quantum Circuits for Long-Range Entangled States}",
    eprint = "2409.07281",
    archivePrefix = "arXiv",
    primaryClass = "quant-ph",
    doi = "10.1103/PhysRevLett.134.170601",
    journal = "Phys. Rev. Lett.",
    volume = "134",
    number = "17",
    pages = "170601",
    year = "2025"
}

@misc{Deshpande:2024kpt,
    author = "Deshpande, Abhinav and Hinsche, Marcel and Najafi, Khadijeh and Sharma, Kunal and Sweke, Ryan and Zoufal, Christa",
    title = "{Dynamic parameterized quantum circuits: expressive and barren-plateau free}",
    eprint = "2411.05760",
    archivePrefix = "arXiv",
    primaryClass = "quant-ph",
    month = "11",
    year = "2024"
}

@article{Farrell:2024mgu,
    author = "Farrell, Roland C. and Illa, Marc and Savage, Martin J.",
    title = "{Steps toward quantum simulations of hadronization and energy loss in dense matter}",
    eprint = "2405.06620",
    archivePrefix = "arXiv",
    primaryClass = "quant-ph",
    reportNumber = "IQuS@UW-21-076, NT@UW-24-06",
    doi = "10.1103/PhysRevC.111.015202",
    journal = "Phys. Rev. C",
    volume = "111",
    number = "1",
    pages = "015202",
    year = "2025"
}

@misc{Barata:2025hgx,
    author = "Barata, Jo{\~a}o and Rico, Enrique",
    title = "{Real-time simulation of jet energy loss and entropy production in high-energy scattering with matter}",
    eprint = "2502.17558",
    archivePrefix = "arXiv",
    primaryClass = "hep-ph",
    reportNumber = "CERN-TH-2025-019",
    month = "2",
    year = "2025"
}

@article{Wang:2020yjh,
    author = "Wang, Samson and Fontana, Enrico and Cerezo, M. and Sharma, Kunal and Sone, Akira and Cincio, Lukasz and Coles, Patrick J.",
    title = "{Noise-Induced Barren Plateaus in Variational Quantum Algorithms}",
    eprint = "2007.14384",
    archivePrefix = "arXiv",
    primaryClass = "quant-ph",
    reportNumber = "LA-UR-20-25526",
    doi = "10.1038/s41467-021-27045-6",
    journal = "Nature Commun.",
    volume = "12",
    pages = "6961",
    year = "2021"
}

@article{Bennewitz:2025nhz,
    author = "Bennewitz, Elizabeth R. and others",
    title = "{Simulating Meson Scattering on Spin Quantum Simulators}",
    eprint = "2403.07061",
    archivePrefix = "arXiv",
    primaryClass = "quant-ph",
    doi = "10.22331/q-2025-06-17-1773",
    journal = "Quantum",
    volume = "9",
    pages = "1773",
    year = "2025"
}

@article{Agrawal:2006cbk,
    author = "Agrawal, Pankaj and Pati, Arun",
    title = "{Perfect teleportation and superdense coding with W states}",
    eprint = "quant-ph/0610001",
    archivePrefix = "arXiv",
    doi = "10.1103/PhysRevA.74.062320",
    journal = "Phys. Rev. A",
    volume = "74",
    number = "6",
    pages = "062320",
    year = "2006"
}

@misc{Yu:2024szp,
    author = "Yu, Jeffery and Muleady, Sean R. and Wang, Yu-Xin and Schine, Nathan and Gorshkov, Alexey V. and Childs, Andrew M.",
    title = "{Efficient preparation of Dicke states}",
    eprint = "2411.03428",
    archivePrefix = "arXiv",
    primaryClass = "quant-ph",
    month = "11",
    year = "2024"
}

@article{Buhrman:2023rft,
    author = "Buhrman, Harry and Folkertsma, Marten and Loff, Bruno and Neumann, Niels M. P.",
    title = "{State preparation by shallow circuits using feed forward}",
    eprint = "2307.14840",
    archivePrefix = "arXiv",
    primaryClass = "quant-ph",
    doi = "10.22331/q-2024-12-09-1552",
    journal = "Quantum",
    volume = "8",
    pages = "1552",
    year = "2024"
}

@misc{Yuan:2025dqo,
    author = "Yuan, Pei and Zhang, Shengyu",
    title = "{Depth-Efficient Quantum Circuit Synthesis for Deterministic Dicke State Preparation}",
    eprint = "2505.15413",
    archivePrefix = "arXiv",
    primaryClass = "quant-ph",
    month = "5",
    year = "2025"
}

@article{Haegeman:2013xcv,
    author = "Haegeman, Jutho and Michalakis, Spyridon and Nachtergaele, Bruno and Osborne, Tobias J. and Schuch, Norbert and Verstraete, Frank",
    title = "{Elementary Excitations in Gapped Quantum Spin Systems}",
    eprint = "1305.2176",
    archivePrefix = "arXiv",
    primaryClass = "quant-ph",
    doi = "10.1103/PhysRevLett.111.080401",
    journal = "Phys. Rev. Lett.",
    volume = "111",
    number = "8",
    pages = "080401",
    year = "2013"
}

@article{Farrell:2023fgd,
    author = "Farrell, Roland C. and Illa, Marc and Ciavarella, Anthony N. and Savage, Martin J.",
    title = "{Scalable Circuits for Preparing Ground States on Digital Quantum Computers: The Schwinger Model Vacuum on 100 Qubits}",
    eprint = "2308.04481",
    archivePrefix = "arXiv",
    primaryClass = "quant-ph",
    reportNumber = "IQuS@UW-21-060, NT@UW-23-13",
    doi = "10.1103/PRXQuantum.5.020315",
    journal = "PRX Quantum",
    volume = "5",
    number = "2",
    pages = "020315",
    year = "2024"
}

@article{VanDyke:2022ffj,
    author = "Van Dyke, John S. and Shirali, Karunya and Barron, George S. and Mayhall, Nicholas J. and Barnes, Edwin and Economou, Sophia E.",
    title = "{Scaling adaptive quantum simulation algorithms via operator pool tiling}",
    eprint = "2206.14215",
    archivePrefix = "arXiv",
    primaryClass = "quant-ph",
    doi = "10.1103/PhysRevResearch.6.L012030",
    journal = "Phys. Rev. Res.",
    volume = "6",
    number = "1",
    pages = "L012030",
    year = "2024"
}

@article{Grimsley:2018wnd,
    author = "Grimsley, Harper R. and Economou, Sophia E. and Barnes, Edwin and Mayhall, Nicholas J.",
    title = "{An adaptive variational algorithm for exact molecular simulations on a quantum computer}",
    eprint = "1812.11173",
    archivePrefix = "arXiv",
    primaryClass = "quant-ph",
    doi = "10.1038/s41467-019-10988-2",
    journal = "Nature Commun.",
    volume = "10",
    pages = "3007",
    year = "2019"
}

@misc{Dalzell:2023ywa,
    author = "Dalzell, Alexander M. and others",
    title = "{Quantum algorithms: A survey of applications and end-to-end complexities}",
    eprint = "2310.03011",
    archivePrefix = "arXiv",
    primaryClass = "quant-ph",
    doi = "10.1017/9781009639651",
    month = "10",
    year = "2023"
}

@inproceedings{Baron:2024kdb,
    author = "Baron, Dror and Patil, Hrushikesh Pramod and Zhou, Huiyang",
    title = "{Qubit-Wise Majority Vote: Maximum Likelihood Quantum Error Mitigation for Algorithms with a Single Correct Output}",
    booktitle = "{2024 International Conference on Quantum Computing and Engineering}",
    eprint = "2402.11830",
    archivePrefix = "arXiv",
    primaryClass = "quant-ph",
    doi = "10.1109/QCE60285.2024.00024",
    month = "2",
    year = "2024"
}

@misc{Chandramouli:2025rfx,
    author = "Chandramouli, Kausthubh and Allen, Kelly Mae and Mori, Christopher and Baron, Dror and Figueiredo, M{\'a}rio A. T.",
    title = "{Statistical Signal Processing for Quantum Error Mitigation}",
    eprint = "2506.00683",
    archivePrefix = "arXiv",
    primaryClass = "quant-ph",
    month = "5",
    year = "2025"
}

@article{Troyer_2005,
	title        = {Computational Complexity and Fundamental Limitations to Fermionic Quantum Monte Carlo Simulations},
	author       = {Troyer, Matthias and Wiese, Uwe-Jens},
	year         = 2005,
	month        = {May},
	journal      = {Phys. Rev. Lett.},
	publisher    = {American Physical Society (APS)},
	volume       = 94,
	number       = 17,
	pages        = 170201,
	doi          = {10.1103/physrevlett.94.170201},
	issn         = {1079-7114},
	url          = {http://dx.doi.org/10.1103/PhysRevLett.94.170201},
	eprint       = {cond-mat/0408370},
	archiveprefix = {arXiv}
}

@article{Benioff1980,
	title        = {The computer as a physical system: A microscopic quantum mechanical Hamiltonian model of computers as represented by Turing machines},
	author       = {Benioff, Paul},
	year         = 1980,
	month        = {May},
	day          = {01},
	journal      = {J. Stat. Phys.},
	volume       = 22,
	number       = 5,
	pages        = {563--591},
	doi          = {10.1007/BF01011339},
	issn         = {1572-9613},
	url          = {https://doi.org/10.1007/BF01011339}
}

@article{Feynman1982,
	title        = {Simulating physics with computers},
	author       = {Feynman, Richard P.},
	year         = 1982,
	journal      = {Int. J. Theor. Phys.},
	volume       = 21,
	pages        = {467--488},
	doi          = {10.1007/BF02650179},
	url          = {https://doi.org/10.1007/BF02650179},
	issue        = 6,
	editor       = {Brown, L. M.}
}

@article{Feynman1986,
	title        = {Quantum mechanical computers},
	author       = {Feynman, Richard P.},
	year         = 1986,
	journal      = {Found. Phys.},
	volume       = 16,
	pages        = {507--531},
	doi          = {10.1007/BF01886518},
	url          = {https://doi.org/10.1007/BF01886518},
	issue        = 6
}

@article{Lloyd1073,
	title        = {Universal Quantum Simulators},
	author       = {Lloyd, Seth},
	year         = 1996,
	journal      = {Science},
	publisher    = {American Association for the Advancement of Science},
	volume       = 273,
	number       = 5278,
	pages        = {1073--1078},
	doi          = {10.1126/science.273.5278.1073},
	issn         = {0036-8075},
	url          = {https://science.sciencemag.org/content/273/5278/1073}
}

@article{Temme:2016vkz,
    author = "Temme, Kristan and Bravyi, Sergey and Gambetta, Jay M.",
    title = "{Error Mitigation for Short-Depth Quantum Circuits}",
    eprint = "1612.02058",
    archivePrefix = "arXiv",
    primaryClass = "quant-ph",
    doi = "10.1103/physrevlett.119.180509",
    journal = "Phys. Rev. Lett.",
    volume = "119",
    number = "18",
    pages = "180509",
    year = "2017"
}

@misc{Wack:2021gvg,
    author = "Wack, Andrew and Paik, Hanhee and Javadi-Abhari, Ali and Jurcevic, Petar and Faro, Ismael and Gambetta, Jay M. and Johnson, Blake R.",
    title = "{Quality, Speed, and Scale: three key attributes to measure the performance of near-term quantum computers}",
    eprint = "2110.14108",
    archivePrefix = "arXiv",
    primaryClass = "quant-ph",
    month = "10",
    year = "2021"
}

@misc{Fischer:2024ipl,
    author = "Fischer, Laurin E. and others",
    title = "{Dynamical simulations of many-body quantum chaos on a quantum computer}",
    eprint = "2411.00765",
    archivePrefix = "arXiv",
    primaryClass = "quant-ph",
    month = "11",
    year = "2024"
}

@article{Deguchi:1999rq,
    author = "Deguchi, Tetsuo and Fabricius, Klaus and McCoy, Barry M.",
    title = "{The sl(2) loop algebra symmetry of the six vertex model at roots of unity. 1. Jordan-Wigner techniques}",
    eprint = "cond-mat/9912141",
    archivePrefix = "arXiv",
    reportNumber = "YITP-SB-99-66",
    doi = "10.1023/A:1004894701900",
    journal = "J. Statist. Phys.",
    volume = "102",
    pages = "701--736",
    year = "2001"
}

@Article{jordan:1928wi,
  author	= "Jordan, Pascual and Wigner, Eugene P.",
  title		= "{About the Pauli exclusion principle}",
  doi		= "10.1007/BF01331938",
  journal	= "Z. Phys.",
  volume	= "47",
  pages		= "631--651",
  year		= "1928"
}

@misc{Haghshenas:2025euj,
    author = "Haghshenas, Reza and others",
    title = "{Digital quantum magnetism at the frontier of classical simulations}",
    eprint = "2503.20870",
    archivePrefix = "arXiv",
    primaryClass = "quant-ph",
    month = "3",
    year = "2025"
}

@misc{Chernyshev:2025lil,
    author = "Chernyshev, Ivan A. and others",
    title = "{Pathfinding Quantum Simulations of Neutrinoless Double-$β$ Decay}",
    eprint = "2506.05757",
    archivePrefix = "arXiv",
    primaryClass = "quant-ph",
    reportNumber = "IQuS@UW-21-103, LA-UR-25-23950",
    month = "6",
    year = "2025"
}

@article{Derby:2020aha,
    author = "Derby, Charles and Klassen, Joel and Bausch, Johannes and Cubitt, Toby",
    title = "{Compact fermion to qubit mappings}",
    eprint = "2003.06939",
    archivePrefix = "arXiv",
    primaryClass = "quant-ph",
    doi = "10.1103/PhysRevB.104.035118",
    journal = "Phys. Rev. B",
    volume = "104",
    number = "3",
    pages = "035118",
    year = "2021"
}

@misc{Constantinides:2025kjx,
    author = "Constantinides, Nathan and Yu, Jeffery and Devulapalli, Dhruv and Fahimniya, Ali and Schaeffer, Luke and Childs, Andrew M. and Gullans, Michael J. and Schuckert, Alexander and Gorshkov, Alexey V.",
    title = "{Low-depth fermion routing without ancillas}",
    eprint = "2510.05099",
    archivePrefix = "arXiv",
    primaryClass = "quant-ph",
    month = "10",
    year = "2025"
}

@misc{Maskara:2025oab,
    author = "Maskara, Nishad and Kalinowski, Marcin and Gonzalez-Cuadra, Daniel and Lukin, Mikhail D.",
    title = "{Fast simulation of fermions with reconfigurable qubits}",
    eprint = "2509.08898",
    archivePrefix = "arXiv",
    primaryClass = "quant-ph",
    month = "9",
    year = "2025"
}

@article{Kivlichan:2018bqq,
    author = "Kivlichan, Ian D. and McClean, Jarrod and Wiebe, Nathan and Gidney, Craig and Aspuru-Guzik, Al{\'a}n and Chan, Garnet Kin-Lic and Babbush, Ryan",
    title = "{Quantum Simulation of Electronic Structure with Linear Depth and Connectivity}",
    eprint = "1711.04789",
    archivePrefix = "arXiv",
    primaryClass = "quant-ph",
    doi = "10.1103/PhysRevLett.120.110501",
    journal = "Phys. Rev. Lett.",
    volume = "120",
    number = "11",
    pages = "110501",
    year = "2018"
}

@misc{Aharonov:2025ssq,
    author = "Aharonov, Dorit and others",
    title = "{Reliable high-accuracy error mitigation for utility-scale quantum circuits}",
    eprint = "2508.10997",
    archivePrefix = "arXiv",
    primaryClass = "quant-ph",
    month = "8",
    year = "2025"
}

@article{Berg:2022ugn,
    author = "Berg, Ewout van den and Minev, Zlatko K. and Kandala, Abhinav and Temme, Kristan",
    title = "{Probabilistic error cancellation with sparse Pauli{\textendash}Lindblad models on noisy quantum processors}",
    eprint = "2201.09866",
    archivePrefix = "arXiv",
    primaryClass = "quant-ph",
    doi = "10.1038/s41567-023-02042-2",
    journal = "Nature Phys.",
    volume = "19",
    number = "8",
    pages = "1116--1121",
    year = "2023"
}

@book{10.1093/acprof:oso/9780198525004.001.0001,
    author = {Giamarchi, Thierry},
    title = {Quantum Physics in One Dimension},
    publisher = {Oxford University Press},
    year = {2003},
    month = {12},
    isbn = {9780198525004},
    doi = {10.1093/acprof:oso/9780198525004.001.0001},
    url = {https://doi.org/10.1093/acprof:oso/9780198525004.001.0001},
}

@article{Robledo-Moreno:2024pzz,
    author = "Robledo-Moreno, Javier and others",
    title = "{Chemistry beyond the scale of exact diagonalization on a quantum-centric supercomputer}",
    eprint = "2405.05068",
    archivePrefix = "arXiv",
    primaryClass = "quant-ph",
    reportNumber = "RIKEN-iTHEMS-Report-24",
    doi = "10.1126/sciadv.adu9991",
    journal = "Sci. Adv.",
    volume = "11",
    number = "25",
    pages = "adu9991",
    year = "2025"
}

@misc{Cobos:2025krn,
    author = {Cobos, Jes{\'u}s and Fraxanet, Joana and Benito, C{\'e}sar and di Marcantonio, Francesco and Rivero, Pedro and Kap{\'a}s, Korn{\'e}l and Werner, Mikl{\'o}s Antal and Legeza, {\"O}rs and Bermudez, Alejandro and Rico, Enrique},
    title = "{Real-Time Dynamics in a (2+1)-D Gauge Theory: The Stringy Nature on a Superconducting Quantum Simulator}",
    eprint = "2507.08088",
    archivePrefix = "arXiv",
    primaryClass = "quant-ph",
    reportNumber = "CERN-TH-2025-111",
    month = "7",
    year = "2025"
}

@article{PhysRevLett.42.673,
  title = {Scaling Theory of Localization: Absence of Quantum Diffusion in Two Dimensions},
  author = {Abrahams, E. and Anderson, P. W. and Licciardello, D. C. and Ramakrishnan, T. V.},
  journal = {Phys. Rev. Lett.},
  volume = {42},
  issue = {10},
  pages = {673--676},
  numpages = {0},
  year = {1979},
  month = {Mar},
  publisher = {American Physical Society},
  doi = {10.1103/PhysRevLett.42.673},
  url = {https://link.aps.org/doi/10.1103/PhysRevLett.42.673}
}

@article{Choi_2016,
   title={Exploring the many-body localization transition in two dimensions},
   volume={352},
   ISSN={1095-9203},
   url={http://dx.doi.org/10.1126/science.aaf8834},
   DOI={10.1126/science.aaf8834},
   number={6293},
   journal={Science},
   publisher={American Association for the Advancement of Science (AAAS)},
   author={Choi, Jae-yoon and Hild, Sebastian and Zeiher, Johannes and Schauß, Peter and Rubio-Abadal, Antonio and Yefsah, Tarik and Khemani, Vedika and Huse, David A. and Bloch, Immanuel and Gross, Christian},
   year={2016},
   month=jun, pages={1547–1552} 
}

@article{Sierant:2024khi,
    author = "Sierant, Piotr and Lewenstein, Maciej and Scardicchio, Antonello and Vidmar, Lev and Zakrzewski, Jakub",
    title = "{Many-body localization in the age of classical computing$^{*}$}",
    eprint = "2403.07111",
    archivePrefix = "arXiv",
    primaryClass = "cond-mat.dis-nn",
    doi = "10.1088/1361-6633/ad9756",
    journal = "Rept. Prog. Phys.",
    volume = "88",
    number = "2",
    pages = "026502",
    year = "2025"
}

@article{PhysRevLett.114.146601,
  title = {Nearest Neighbor Tight Binding Models with an Exact Mobility Edge in One Dimension},
  author = {Ganeshan, Sriram and Pixley, J. H. and Das Sarma, S.},
  journal = {Phys. Rev. Lett.},
  volume = {114},
  issue = {14},
  pages = {146601},
  numpages = {5},
  year = {2015},
  month = {Apr},
  publisher = {American Physical Society},
  doi = {10.1103/PhysRevLett.114.146601},
  url = {https://link.aps.org/doi/10.1103/PhysRevLett.114.146601}
}

@inproceedings{baron2024qubit,
  title={Qubit-wise majority vote: Maximum likelihood quantum error mitigation for algorithms with a single correct output},
  author={Baron, Dror and Patil, Hrushikesh Pramod and Zhou, Huiyang},
  booktitle={2024 IEEE International Conference on Quantum Computing and Engineering (QCE)},
  volume={1},
  pages={124--133},
  year={2024},
  organization={IEEE}
}

@article{Cruz_2019,
    author={Cruz, Diogo and Fournier, Romain and Gremion, Fabien and Jeannerot, Alix and Komagata, Kenichi and Tosic, Tara and Thiesbrummel, Jarla and Chan, Chun Lam and Macris, Nicolas and Dupertuis, Marc‐André and Javerzac‐Galy, Clément},
   title= "{Efficient Quantum Algorithms for GHZ and W States, and Implementation on the IBM Quantum Computer}",
   eprint = "1807.05572",
   archivePrefix = "arXiv",
   primaryClass = "quant-ph",  
   journal= "Adv. Quantum Technol.",
   volume= "2",
   doi = "10.1002/qute.201900015",
   number= "5-6",
    pages= "1900015",
   year= "2019"
}

@article{Dur:2000zz,
    author = "Dur, W. and Vidal, G. and Cirac, J. I.",
    title = "{Three qubits can be entangled in two inequivalent ways}",
    eprint = "quant-ph/0005115",
    archivePrefix = "arXiv",
    doi = "10.1103/PhysRevA.62.062314",
    journal = "Phys. Rev. A",
    volume = "62",
    pages = "062314",
    year = "2000"
}

@misc{Farrell:2025nkx,
    author = "Farrell, Roland C. and Zemlevskiy, Nikita A. and Illa, Marc and Preskill, John",
    title = "{Digital quantum simulations of scattering in quantum field theories using W states}",
    eprint = "2505.03111",
    archivePrefix = "arXiv",
    primaryClass = "quant-ph",
    reportNumber = "IQuS@UW-21-099",
    month = "5",
    year = "2025"
}

@article{Yu:2022ivm,
    author = "Yu, Hongye and Zhao, Yusheng and Wei, Tzu-Chieh",
    title = "{Simulating large-size quantum spin chains on cloud-based superconducting quantum computers}",
    eprint = "2207.09994",
    archivePrefix = "arXiv",
    primaryClass = "quant-ph",
    doi = "10.1103/PhysRevResearch.5.013183",
    journal = "Phys. Rev. Res.",
    volume = "5",
    number = "1",
    pages = "013183",
    year = "2023"
}

@article{Piroli:2024mcm,
  title={Approximating many-body quantum states with quantum circuits and measurements},
  author={Piroli, Lorenzo and Styliaris, Georgios and Cirac, J Ignacio},
  journal={Physical Review Letters},
  volume={133},
  number={23},
  pages={230401},
  year={2024},
  publisher={APS}
}

@inproceedings{Bartschi:2022dicke,
  title={Short-depth circuits for {D}icke state preparation},
  author={B{\"a}rtschi, Andreas and Eidenbenz, Stephan},
  booktitle={2022 IEEE International Conference on Quantum Computing and Engineering (QCE)},
  pages={87--96},
  year={2022},
  organization={IEEE}
}

\end{document}